\documentclass[a4paper,fleqn,usenatbib]{mnras}

\usepackage{newtxtext,newtxmath}

\usepackage[T1]{fontenc}
\usepackage{ae,aecompl}

\usepackage{graphicx}	
\usepackage{amsmath}	
\usepackage{amssymb}	

\usepackage{booktabs}

\title[Temperature and Bolometric Luminosity Evolution in SNe-II]{The Evolution of Temperature and Bolometric Luminosity in Type-II Supernovae}

\author[Faran et al.]{
T. Faran,$^{1}$\thanks{E-mail: \href{mailto:tamar.faran@mail.huji.ac.il}{tamarfaran@mail.huji.ac.il}}
E. Nakar,$^{2}$
and D. Poznanski$^{2}$
\\
$^{1}$Racah Institute of Physics, The Hebrew University of Jerusalem, Jerusalem 91904, Israel \\
$^{2}$School of Physics and Astronomy, Tel-Aviv University, Tel Aviv 69978, Israel.\\	
}

\date{Accepted XXX. Received YYY; in original form ZZZ}

\pubyear{2017}

\begin{document}
\label{firstpage}
\pagerange{\pageref{firstpage}--\pageref{lastpage}}
\maketitle

\begin{abstract}
In this work we present a uniform analysis of the temperature evolution and bolometric luminosity of a sample of 29 type-II supernovae (SNe), by fitting a black body model to their multi-band photometry. Our sample includes only SNe with high quality multi-band data and relatively well sampled time coverage. Most of the SNe in our sample were detected less than a week after explosion so their light curves cover the evolution both before and after recombination starts playing a role. We use this sample to study the signature of hydrogen recombination, which is expected to appear once the observed temperature drops to $\approx 7,000$K. Theory predicts that before recombination starts affecting the light curve, both the luminosity and the temperature should drop relatively fast, following a power-law in time. Once the recombination front reaches inner parts of the outflow, it sets the observed temperature to be nearly constant, and slows the decline of the luminosity (or even leads to a re-brightening). We compare our data to analytic studies and find strong evidence for the signature of recombination. We also find that the onset of the optical plateau in a given filter, is effectively the time at which the black body peak reaches the central wavelength of the filter, as it cools, and it does not correspond to the time at which recombination starts affecting the emission.
\end{abstract}



\section{Introduction}

\label{s:intro}
Type II supernovae (SNe) are defined by the prominent hydrogen lines in their spectra. They are believed to originate from the collapse of an iron core of massive stars ($\gtrsim$8 $M_{\sun}$) that retain their hydrogen envelope. The most common sub-type, comprising $\sim 70$ percent of all type II SNe, is characterized by a phase of of a roughly constant magnitude in the optical bands, hence their name type II-Plateau (II-P). This plateau phase typically starts 1--2 weeks after the explosion and lasts for $\sim$100\,d. Pre-explosion images have revealed that the progenitors of this class are red supergiants, in the mass range of (7--16 $M_{\sun}$) (\citealt{Smartt2015}; for individual progenitor detections see e.g. \citealt{VanDyk2003a}, \citealt{VanDyk2003b}, \citealt{VanDyk2012}). Type II-Linear (II-L) SNe constitutes another subclass of type II SNe \citep[e.g.,][]{Patat1994, Arcavi2012, Faran2014a, Faran2014b}. They are spectroscopically very similar to type II-P events \citep[see]{Faran2014b}, but their light curves are declining in all bands. In both types (II-P and II-L) there is typically a sharp drop in the luminosity after $\sim$100\,d and the luminosity starts to follow roughly the exponential decay expected from emission powered by the decay of $^{56}$Ni. The distinction between these two classes is not well defined, and studies have used different definitions for II-L SNe. However, several recent works have shown that there exists a continuum of decline rates between slow declining and fast declining SNe \citep{Anderson2014, Faran2014a}, which suggests that a separation into two different classes may be artificial. Other type II sub-classes will not be discussed here. In this paper we focus on the light curves of type II-P and type II-L SNe, without making the distinction between the two types, and refer to them in short as type II SNe. Our goal here is to perform a uniform analysis of the bolometric luminosity and temperature evolution of a large sample of type II SNe and to compare our findings to theoretical models, focusing on the transition to the plateau, which takes place during the first two weeks.

There are several dozen type II SNe with detailed multi-wavelength observations. These are typically presented and analyzed individually \citep[e.g.,][]{Leonard2002b,Maguire2010b,Pastorello2009,Inserra2011,Takats2014,Takats2015,Fraser2011,Tomasella2013,DallOra2014, Barbarino2015, Valenti2015}. There are only a few studies that analyze the bolometric properties and temperatures of a sample of type-II SNe. \citet{Bersten2009} extracted bolometric light curves and effective temperature evolution for 33 SNe II-P, using calibrations for bolometric corrections from 3 well observed SNe. \citet{Valenti2016} derived the effective temperature from black body fits to photometric data of 30 type-II SNe, and calculated pseudo-bolometric light curves by integration over the optical bands. \citet{Lusk2016} provides bolometric light curves and temperatures for 5 peculiar type II-P SNe that originated from blue supergiants, by integrating over the observed photometry and correcting for the missing flux in the UV.

The theoretical interpretation of type II light curves is that the emission until the end of the plateau is dominated by the cooling emission, i.e., the leakage of radiation energy that was deposited in the envelope by the shock that unbinds it \citep{FalkArnett1977}. Energy deposited by the decay of $^{56}$Ni may contribute to this phase \citep[e.g.,][]{FalkArnett1977,Young2004, Utrobin2007}, but this contribution is found to be subdominant \citep{Nakar2016}. The end of the plateau marks the release of all the internal energy deposited by the shock in the envelope. At later times the SN enters its nebular phase and the entire luminosity is driven by the decay of $^{56}$Ni.  

During the early stages of the light curve ($\sim 1-3$ weeks) the leakage of radiation is facilitated mostly by the drop in the optical depth of the outflow due to its expansion \citep{Arnett1980}. Models predict that during this phase both the temperature and the bolometric luminosity drop roughly as power-laws in time \citep{NakarSari2010,Rabinak2011,Shussman2016b}. Once the observed temperature drop to $\approx 7000$\,K, hydrogen recombination becomes important and a recombination front starts moving from the outside towards inner parts of the ejecta. During this phase, recombination, rather then expansion, is the main driver of the drop in the optical depth of the outflow. As a result the observed temperature remains almost constant while the luminosity starts dropping much more slowly or even rises.

In this work we derive the temperatures and bolometric evolution of 29 type-II SNe with high quality multi-band light curves, by fitting a black body spectrum to their spectral energy distribution (SEDs). As will be discussed extensively below, these are far from trivial. SNe II are not blackbodies, and at different times several effects lead to systematic offsets from a pure black body in various bands. Nevertheless, guided by the data and theoretical insight, we derive the underlying black body properties. We compare the temporal evolution of the temperature and luminosity to theoretical predictions, paying special attention to signs of the recombination processes in the envelope. In Section \ref{s:sample} we describe the contents of our SN sample and the data, in Section \ref{s:fitting} we explain how the  black body fits to the data are done. Section \ref{s:results} describes the results of the fitting and the possible effect of extinction on the results, and Section \ref{s:theory} presents a comparison of our results to theoretical expectations. We summarize our results in Section. \ref{s:Summary}.

\section{The Sample}
\label{s:sample}
We construct from the literature a sample of 29 type-II SNe with good temporal coverage and multi-band photometry. The sample mostly relies on the SNe collected in \citet{Pejcha2015}, \citet{Faran2014a} and \citep{Faran2014b}. The SNe in the sample were required to have sufficiently early data (starting less than 20 days after the explosion) and a well sampled light curves so the early behavior could be compared to the late behavior and to theoretical models. Some objects did not enter the sample despite having well sampled photometric curves, because their data was not good enough to produce good quality temperature and luminosity curves. Ten of the objects have \textit{Swift} UV observations, 10 have JHK data, and 7 objects have both JHK and UV. The photometric data were corrected for galactic extinction according to \citet{Cardelli1989}, but not for host galactic extinction, since there is no method that can provide an accurate estimate for $E(B-V)_{host}$ (see for example the discussion in \citealt{Faran2014a}). We note however, that \citet{Faran2014a} found that $E(B-V)_{host}$ is typically small, of order 0.1 for nearby SNe. The explosion day is set as the mid-point between the first detection and the last non-detection of the SN, and the uncertainty is conservatively set as half the difference. Distance measurements were collected from NED\footnote{The NASA/IPAC Extragalactic Database (NED) is operated by the Jet Propulsion Laboratory, California Institute of Technology, under contract with the National Aeronautics and Space Administration (NASA).} and averaged, using only distances based on the Tully-Fisher method, Cepheids, and SNe~Ia. All of the objects are at low redshift with z$<$0.03. The SN properties and their references are summarized in Table. \ref{t:sample}.

\begin{table*}
\centering
\caption{SN Sample Details}
\centering
\begin{tabular}{llcccl}
\hline
SN name & Bands & Explosion day (MJD) & $\mu$ & z$_{host}$ & Reference \\ 
\hline
SN1999em & U,B,V,R,I,J,H,K & 51476 $\pm$ 4 & 29.84 $\pm$ 0.05 & 0.002 & \citet{Leonard2002} \\ 
&&&&&\citet{Pejcha2015}\\ 
SN1999gi & B,V,R,I & 51519 $\pm$ 4 & 30.24 $\pm$ 0.04 & 0.002 & \citet{Faran2014a} \\ 
SN2000dc & B,V,R,I & 51762 $\pm$ 4 & 32.93 $\pm$ 0.14 & 0.010 & \citet{Faran2014a} \\ 
SN2001cm & B,V,I & 52064 $\pm$ 1 & 33.18 $\pm$ 0.10 & 0.011 & \citet{Faran2014a} \\ 
SN2001cy & B,V,R,I & 52086 $\pm$ 6 & 33.01 $\pm$ 0.12 & 0.015 & \citet{Faran2014b} \\ 
SN2001do & B,V,R,I & 52134 $\pm$ 2 & 32.35 $\pm$ 0.15 & 0.010 & \citet{Faran2014b} \\ 
SN2001fa & B,V,R,I & 52198 $\pm$ 3 & 34.24 $\pm$ 3.42 & 0.017 & \citet{Faran2014b} \\ 
SN2001x & B,V,R,I & 51963 $\pm$ 5 & 31.59 $\pm$ 0.11 & 0.005 & \citet{Faran2014a} \\ 
SN2002gd & B,V,R,I & 52553 $\pm$ 15 & 32.90 $\pm$ 0.21 & 0.009 & \citet{Faran2014a} \\ 
SN2003hf & B,V,R,I & 52864 $\pm$ 2 & 35.51 $\pm$ 3.55 & 0.031 & \citet{Faran2014b} \\ 
SN2003hk & B,V,R,I & 52860 $\pm$ 2 & 34.41 $\pm$ 0.20 & 0.023 & \citet{Faran2014b} \\ 
SN2003iq & B,V,R,I & 52920 $\pm$ 2 & 32.28 $\pm$ 0.08 & 0.008 & \citet{Faran2014a} \\ 
SN2003z & B,V,R,I & 52665 $\pm$ 5 & 31.23 $\pm$ 3.12 & 0.004 & \citet{Faran2014a} \\ 
SN2004A & B,V,R,I & 53007 $\pm$ 7 & 31.61 $\pm$ 0.32 & 0.003 & \citet{Gurugubelli2008}\\
&&&&&\citet{Maguire2010b}\\ 
SN2004du & B,V,R,I & 53228 $\pm$ 2 & 33.94 $\pm$ 0.13 & 0.017 & \citet{Faran2014a} \\ 
SN2004et & U,B,V,R,I,J,H,K & 53271 $\pm$ 1 & 28.41 $\pm$ 0.07 & 0.000 & \citet{Maguire2010b} \\ 
SN2005cs & Swift UVOT,U,B,V,R,I,J,H,K & 53549 $\pm$ 0 & 29.36 $\pm$ 0.01 & 0.002 & \citet{Pastorello2009} \\ 
SN2006bp & Swift UVOT,U,B,V,r,i & 53834 $\pm$ 1 & 31.11 $\pm$ 0.05 & 0.004 & \citet{Quimby2007} \\ 
SN2007od & Swift UVOT,U,B,V,R,I,J,H,K & 54399 $\pm$ 8 & 32.29 $\pm$ 0.17 & 0.006 & \citet{Inserra2011} \\ 
SN2008in & Swift UVOT,U,B,V,R,I,J,H & 54822 $\pm$ 10 & 30.52 $\pm$ 0.09 & 0.005 & \citet{Roy2012} \\ 
SN2009N & Swift UVOT,B,g,V,R,r,I,i,J,H & 54845 $\pm$ 11 & 31.68 $\pm$ 0.08 & 0.003 & \citet{Takats2014} \\ 
SN2009bw & Swift UV,U,B,V,R,I,J,H,K & 54917 $\pm$ 3 & 30.60 $\pm$ 0.02 & 0.004 & \citet{Inserra2012a} \\ 
SN2009ib & U,u,B,g,V,R,r,I,i,J,H & 55041 $\pm$ 10 & 31.48 $\pm$ 0.31 & 0.004 & \citet{Takats2015} \\ 
SN2012A & Swift UVOT,U,B,g,V,R,r,I,i,J,H,K & 55929 $\pm$ 5 & 29.72 $\pm$ 0.17 & 0.003 & \citet{Tomasella2013} \\ 
SN2012aw & Swift UVOT,U,u,B,g,V,R,r,I,i,J,H,K & 56002 $\pm$ 1 & 29.89 $\pm$ 0.07 & 0.003 & \citet{Bose2013}\\
&&&&&\citet{DallOra2014} \\ 
SN2012ec & u,B,g,V,R,i,J,H,K & 56143 $\pm$ 10 & 31.57 $\pm$ 0.45 & 0.005 & \citet{Barbarino2015} \\ 
SN2013ab & Swift UVOT,U,B,g,V,R,r,I,i & 56340 $\pm$ 1 & 31.71 $\pm$ 0.66 & 0.005 & \citet{Bose2015a} \\ 
SN2013by & Swift UV,u,B,g,V,r,i & 56407 $\pm$ 11 & 30.84 $\pm$ 0.15 & 0.004 & \citet{Valenti2015} \\ 
SN2013ej & Swift UV *,U,u,B,g,V,R,r,I,i & 56497 $\pm$ 1 & 29.77 $\pm$ 2.98 & 0.002 & \citet{Richmond2014}\\
&&&&&\citet{Valenti2014} \\ 
\hline
\end{tabular}
\label{t:sample}
\end{table*}

\section{Black Body Fitting}
\label{s:fitting}

We calculate the temperature and bolometric luminosity of the SNe at each epoch by fitting a black-body to the photometric data, according to $\rm L_{bol}=4\pi\sigma T^4R^2$. We create a two-dimensional grid of temperatures evenly spaced by 20K, and black body radii (R) in the range of 10$^{12}$--10$^{16}$cm with spacing that corresponds to 0.002 mag. We then compute synthetic photometry from the black body distribution for every T and R values in each of the filter bands. Since data in different photometric bands were sometimes taken at different epochs, linear interpolation is used to account for the missing epochs. The interpolation is constrained to a maximum of 10 days from the nearest data point at early or late phases (before day 10 or after day 70), and to 20 days during intermediate phases, where the SN properties evolve more slowly.

A correct estimation of the photometric uncertainties is needed when fitting a black body to the photometry. Due to the relatively small number of data points, the fit is sensitive to errors that are under- or over-estimated. We therefore set a minimum value of 0.05 magnitudes to the error (such that the error is the maximal value between the given photometric error and 0.05 mag), which is a typical value for the scatter in our light curves. We assign the effective wavelength of the filter transmission curve to each band, and fit the data to find the black body temperature and radius by minimizing $\rm \chi^{2}$. The uncertainty on the temperature is found by marginalizing the likelihood over the radius and finding the upper and lower temperature where $\rm \chi^{2}=\chi_{min}^{2}+1$. To find the uncertainty in the luminosity, we calculate L$_{bol}$ for every T and R, and find the contour in which $\rm \chi^{2}=\chi_{min}^{2}+1$. The maximal and minimal values of L$_{bol}$ are taken to be the upper and lower errors, respectively.

The SN spectrum is expected to follow a black body shape only in a limited frequency range, where $\rm h\nu\sim kT$. At high frequencies the flux is suppressed by line blanketing, and at much lower frequencies, in the Rayleigh-Jeans (RJ) regime, it is predicted to be brighter than the RJ tail due to the fact that the thermalization depth in this range is frequency dependent \citep{Shussman2016b}. We observe both effects in our data, and fit a black body only to the wavelength regions where it provides a good approximation.

In agreement with the theoretical predictions, we see that at high temperatures JHK observations cannot be well described by a standard black body spectrum, and tend to systematically lie above the RJ tail. This effect was recently modeled analytically by \citet{Shussman2016b}  and will be further discussed in section \ref{s:deviation_from_BB}. In the cases where this discrepancy is observed, we use only UV and optical data, and exclude the JHK bands.

\begin{figure}
 \centering
\includegraphics[width=0.5\textwidth]{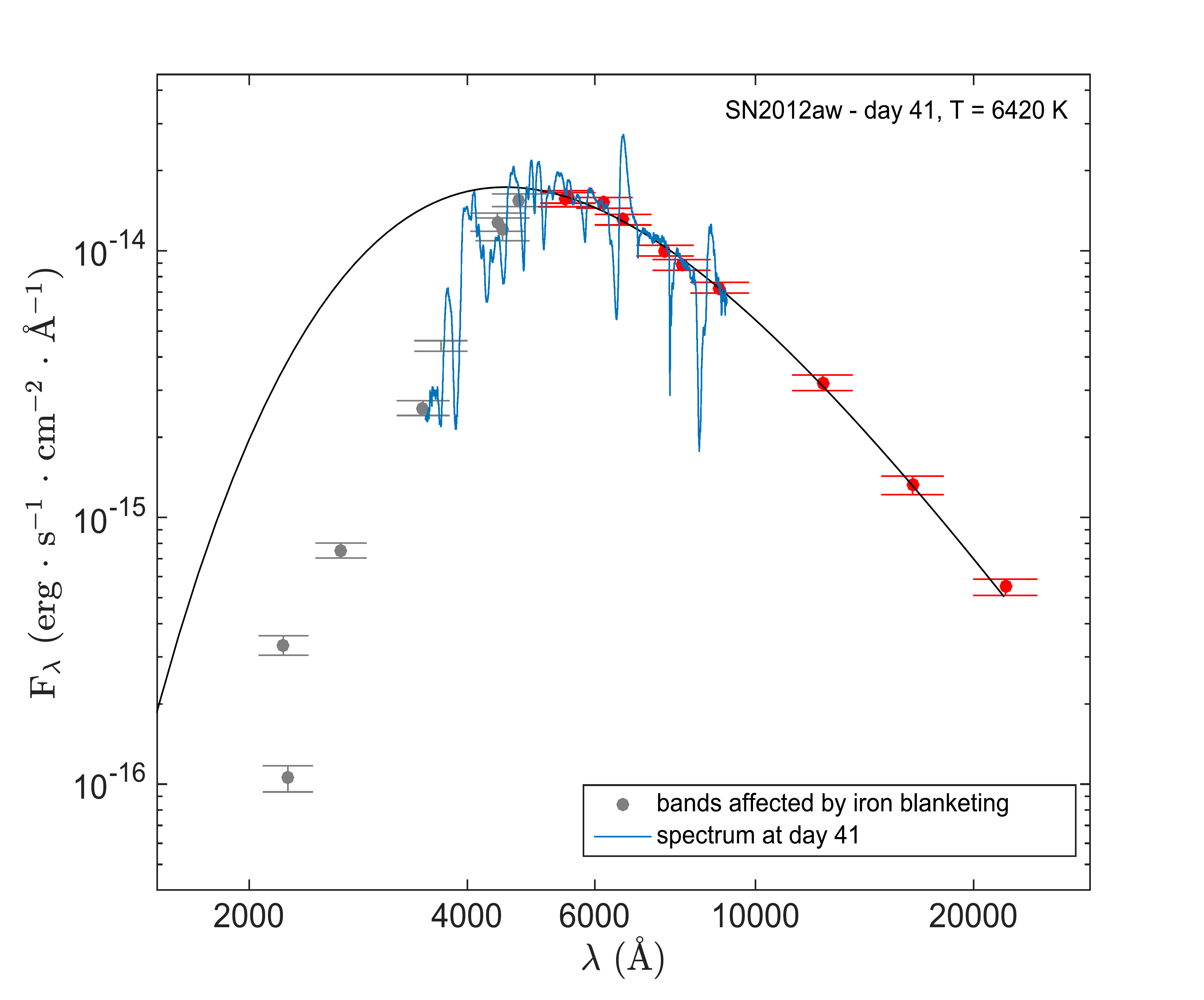} 
\caption{Black body fit of SN2012aw on day 41. The grey points represent bands that are affected by iron-blanketing and were therefore excluded from the fit. The black body curve fits the data very well above 5000$\rm \AA$, whereas at shorter wavelengths the SED is no longer represented by a black body. A spectrum taken on day 41 \citep{Bose2013} is also shown to coincide well with the data, and confirms the flux cutoff around the $B$-band.}\label{f:line_blanketing}
 \end{figure}

As the temperature drops below $\sim 10,000-12,000$K, line blanketing by iron group elements becomes strong and creates a deficiency in the measured UV flux, compared to a pure black body. The main species responsible for the strong absorption are Fe III and Ti III \citep{Kasen2009}. Line opacity is highly sensitive to the temperature, and even a slight cooling of the photosphere induces a fast recombination of Fe III and Ti III to Fe II and Ti II \citep{Kasen2009,Eastman1996}. The flux absorption becomes stronger and shifts further to the optical bands as the temperature continues to decrease to $~8,000$K. Figure.\ref{f:line_blanketing} demonstrates the effect of line blanketing on the SED of SN2012aw on day 41. Data taken at wavelengths shorter than ~5000$\rm\AA$ were found to be affected by line blanketing and were excluded from the fit (grey points), and only bands with wavelengths longer than 5000$\rm\AA$ were used (red points). The resulting black body at a temperature of 6420K fits the red points very well and is also in very good agreement with a spectrum taken at the same epoch. The observed spectrum also confirms the flux cut-off around the $B$-band. In order to determine the time where the flux in each photometric band is suppressed by line-blanketing, we run the black body fitting procedure on each of the following filter groups: UV-UBVRIJHK, UBVRIJHK, BVRIJHK, VRIJHK and RIJHK, i.e., each time excluding the bluest band. We first determine, as an example, when the flux in the UV bands falls below the black body curve by looking at the fit to the UBVRIJHK regime. As long as the temperature is high enough, the UV flux will appear above the black body fit to UBVRIJHK or right on it. However, as the temperature decreases enough such that line blanketing starts to have an effect on the UV flux, the UV data points will drop below the UBVRIJHK curve. We exclude a certain band from the fit when it is 1-$\sigma$ below the black body curve. This means that until that epoch we can use the UV-UBVRIJHK bands to determine the black body parameters, and from that day on we can only use the UBVRIJHK bands to fit the data. This procedure is repeated with the other bands to determine when the U, B, and V bands are affected by line blanketing and need to be excluded. The transition days coincide with the intersection between the temperature curves calculated with the bluest band, and the one calculated without it. Eventually, we construct the final temperature and luminosity curves using the transitions determined for each of the objects.

At early phases, when the temperature is higher than 10$^{4}$K, the peak of $\rm F_{\lambda}$ occurs at wavelengths shorter than 3000$\rm \AA$ and UV observations are therefore crucial to constrain the fit parameters. JHK observations lie far from the peak of $\rm F_{\lambda}$ even at low temperatures ($\sim$6000\,K), and therefore do not play a critical role in constraining the temperature. However, due to the exclusion of many of the bluer bands by line blanketing, it is necessary to have more data points in the red to improve the fit, meaning that JHK data become important at late epochs.

\begin{figure}
 \centering
\includegraphics[width=0.5\textwidth]{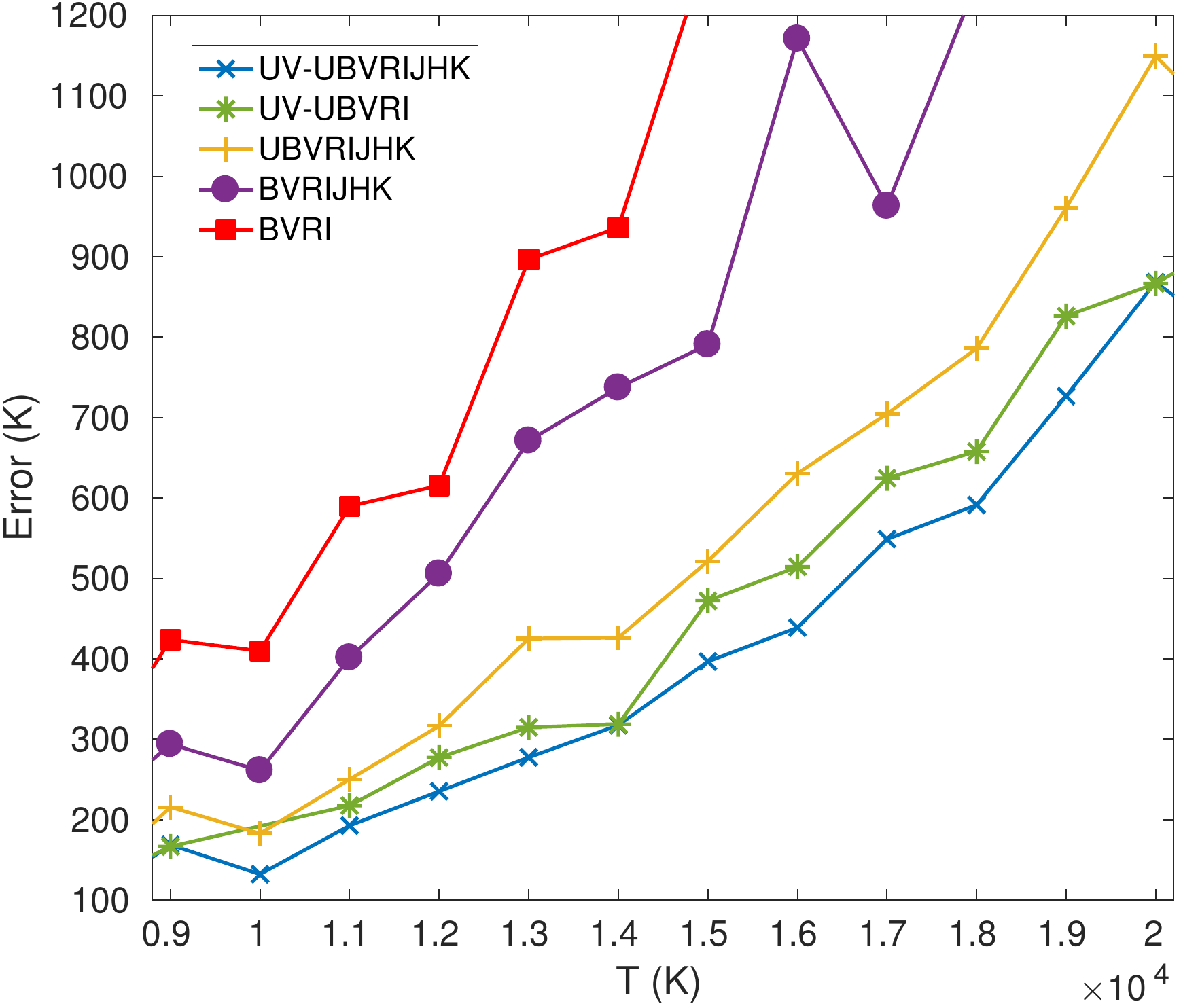} 
\includegraphics[width=0.5\textwidth]{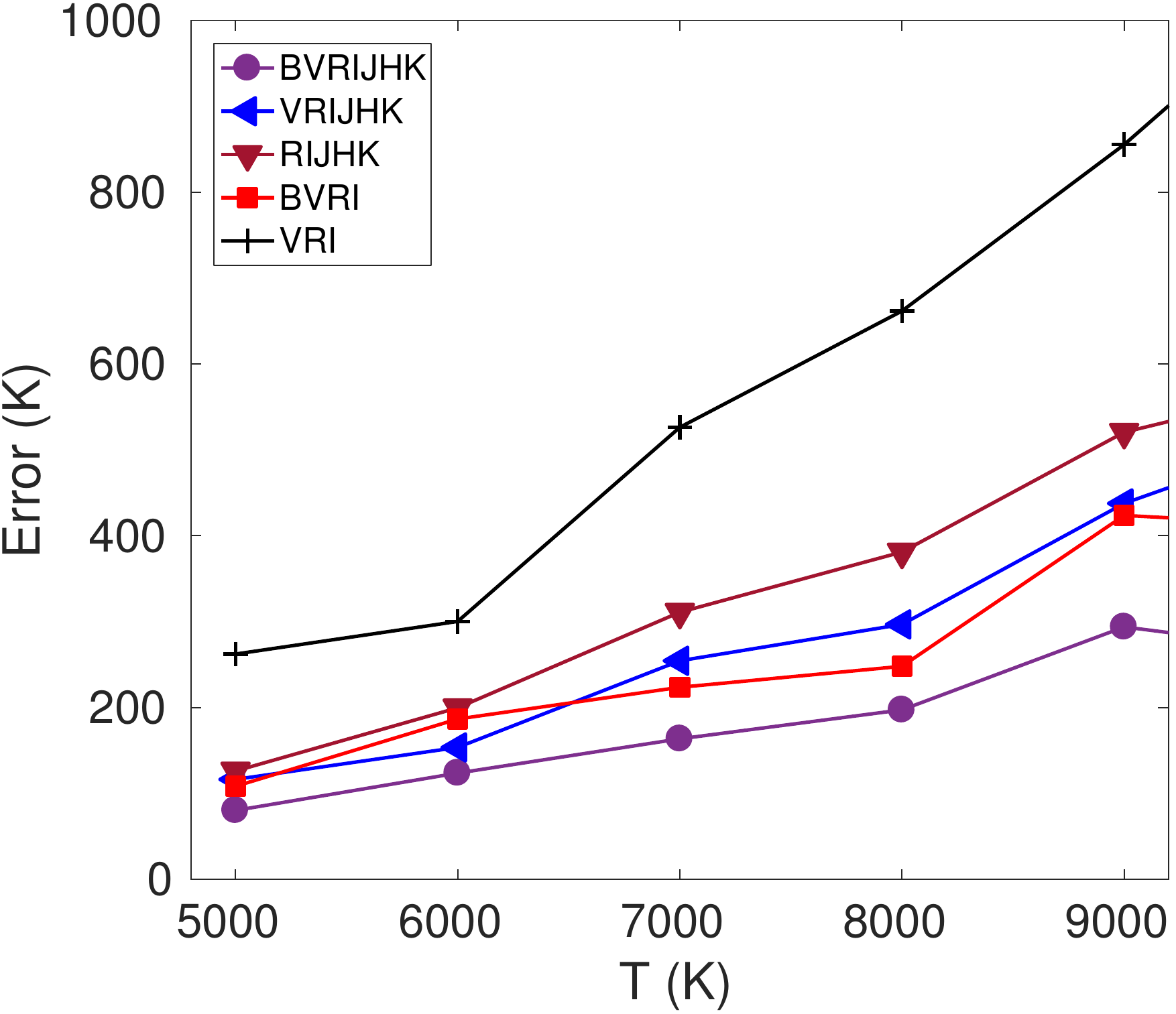} 
\caption{The expected uncertainties of the temperature of a black body, resulting from a simulation of synthetic data. The estimated uncertainties decrease as a function of the proximity of the bands to the black body peak, and therefore rise with the temperature. }\label{f:simulation_errors}
 \end{figure}

In order to quantify the importance of UV and Infra-Red (IR) photometry, we run a simulation and estimate the expected errors on the temperature in the absence of UV and IR. We produce synthetic photometry from black body distributions at temperatures 5000K--25,000K in 1000K bins, simulating a spread in the data using the typical photometric errors in each band. We then fit the synthetic data to a black body, repeating the process 100 times per temperature bin. The mean value and standard deviation (STD) of the best-fitting temperatures are computed, where we treat the STD as a measure of the typical statistical error. The uncertainties deduced from the simulation are presented as a function of the temperature in Figure. \ref{f:simulation_errors}. From the upper panel of Figure. \ref{f:simulation_errors} one can see that at temperatures of $\sim$20,000K, the uncertainties on the temperature are quite high (over 800K) even with UV data. This reflects the fact that $\rm F_{\lambda}$ peaks at $\sim$1300$\rm \AA$, while the effective wavelength of the bluest filter we use ($Swift-uvw2$) is only at 2230$\rm\AA$. Below T=15,000K, fits that do not include UV (but do include U) are able to reproduce the temperature with an accuracy of $\sim$500K. When $U$-band data is excluded, the fit reaches that accuracy below T=12,000K.

At lower temperatures, corresponding to late epochs, most of the flux at wavelengths shorter than the B band is already affected by line blanketing and only bands with effective wavelengths longer than the V band can be used. In the bottom panel of Figure \ref{f:simulation_errors} it is evident that JHK data are important at T$>$7000K if the B band is not included, as the STD of the fit temperature is relatively high and rises rapidly with the model temperature. Although fitting with the V, R and I bands is still able to produce errors below 10\%, we will see in Section \ref{s:Temperature} that the flux in the V band is typically absorbed by iron blanketing at $\sim$6000K. In the absence of JHK observations, we are left with only 3 data points for many objects -  V, R and I. In these cases, it is impossible to determine when the V band falls below the black body curve, since we cannot examine the fit done without V, having only 2 data points at longer wavelengths. As a result, when an object does not have JHK data we typically cannot trust the temperature curve below $\sim$6000K and we do not fit the data below this temperature.

Throughout this paper, we consider only objects with U or UV data to deduce physical parameters at high temperatures (above $\sim$10,000K), and objects with JHK data at low temperatures (below $\sim$6000K).

\section{Results}
\label{s:results}
\subsection{Temperature}
\label{s:Temperature}

The temperature curves are computed from the black body fits at each epoch and are presented in Figure. \ref{f:T} (a list of all the results is also available in Appendix. \ref{s:results}). After the explosion, the envelope expands and cools adiabatically. The typical temperatures during the first 10 days are above 10,000 degrees. In cases where UV data exist, the typical errors for that temperature range are smaller than $\sim$500K, and are comparable to the errors predicted by our simulations (see Section \ref{s:fitting}). Between 20 and 40 days after the explosion, the temperature curves start evolving more slowly compared to early phases. The flattening typically happens between 6000K and 7000K, and is therefore consistent with being associated with a recombination wave that propagates into the envelope and dictates the black body temperature to be the temperature of hydrogen recombination. This effect is analyzed and discussed in Section \ref{s:logarithmic_deriv}. We note that objects without JHK data do not show this flattening, since as discussed in Section \ref{s:fitting}, for these SNe we were not able to determine the time where the $V$-band can no longer be used and the fit stops when the temperature reaches 6000K.

As in \citet{Valenti2016}, we observe that excluding UV data from the fit systematically leads to lower temperatures. This is true also for U, B and V, where the temperature produced without the bluest band is lower than that produced with it, before its flux is affected by iron blanketing. 
Since we do not observe this behavior in the simulation described in Section \ref{s:fitting}, the effect is not statistical and points to a deviation of the spectrum from a black body. We suggest that this is related to the re-distribution of energy that is absorbed by line blanketing. Most of the absorbed radiation is expected to be re-emitted close to the absorption wavelength \citep[see][]{Pinto2000}. As a result, the flux of the bluest band we use will be higher than the black body at the same temperature. Since the bands near the peak of the spectrum have the highest effect on the fit, that will result in higher fit temperatures. We redo the fits without the bluest band, and measure the flux under the resulting black body curve. We then measure the flux excess in the bands that lie above the black body curve, and the flux deficiency (due to line blanketing) in the bands below the black body curve and find that they are of the same order. This reinforces the assumption that the absorbed radiation by iron group elements is emitted at wavelengths close to the black body peak, and may add an uncertainty to the temperature and bolometric luminsoity that we measure. Above 10,000K, the temperatures calculated with the bluest band are $\approx 10 \%$ higher than the ones calculated when it is omitted, and the difference becomes less significant at lower temperatures. The effect on the luminosity is higher and can get up to $\approx 10-20 \%$. Therefore, the temperatures and bolometric luminosities presented in this paper can be overestimated by up to $\approx 10-20 \%$.

We record the temperatures at which the flux in different bands starts being affected by line blanketing, and find that the typical temperature for UV is $\sim$11,000K, and $\sim$8000K in the $U$ and $B$ bands. The $V$ band seems to be affected around $\sim$6000K. These results agree with the temperatures shown in \citet{Eastman1996}'s Figure. 7.

\citet{Bersten2009} fit a black body to the photometry of SN1999em corrected to $\rm A_{V}^{host}=0.18$ and present its temperature and bolometric luminosity curves. After correcting our data to $\rm A_{V}^{host}=0.18$, we extract the temperature curve and compare it to the middle panel in \citet{Bersten2009}'s Figure. 8. We find a good agreement between the values of the temperature and its evolution. The temperature computed at the first epoch, $\sim5$ days, is around 13,000K in both curves and decreases to show a "bump" around day 16. Eventually, both curves settle on a temperature of $\rm \sim6000K$ in the middle of the plateau. \citet{Valenti2016} also fit a black body to several SNe that are included in our sample, but unfortunately the values are not provided, and we cannot perform a quantitative comparison.

\subsection{Luminosity}
\label{s:Luminosity}

\begin{figure*}
 \centering
\includegraphics[width=0.85\textwidth]{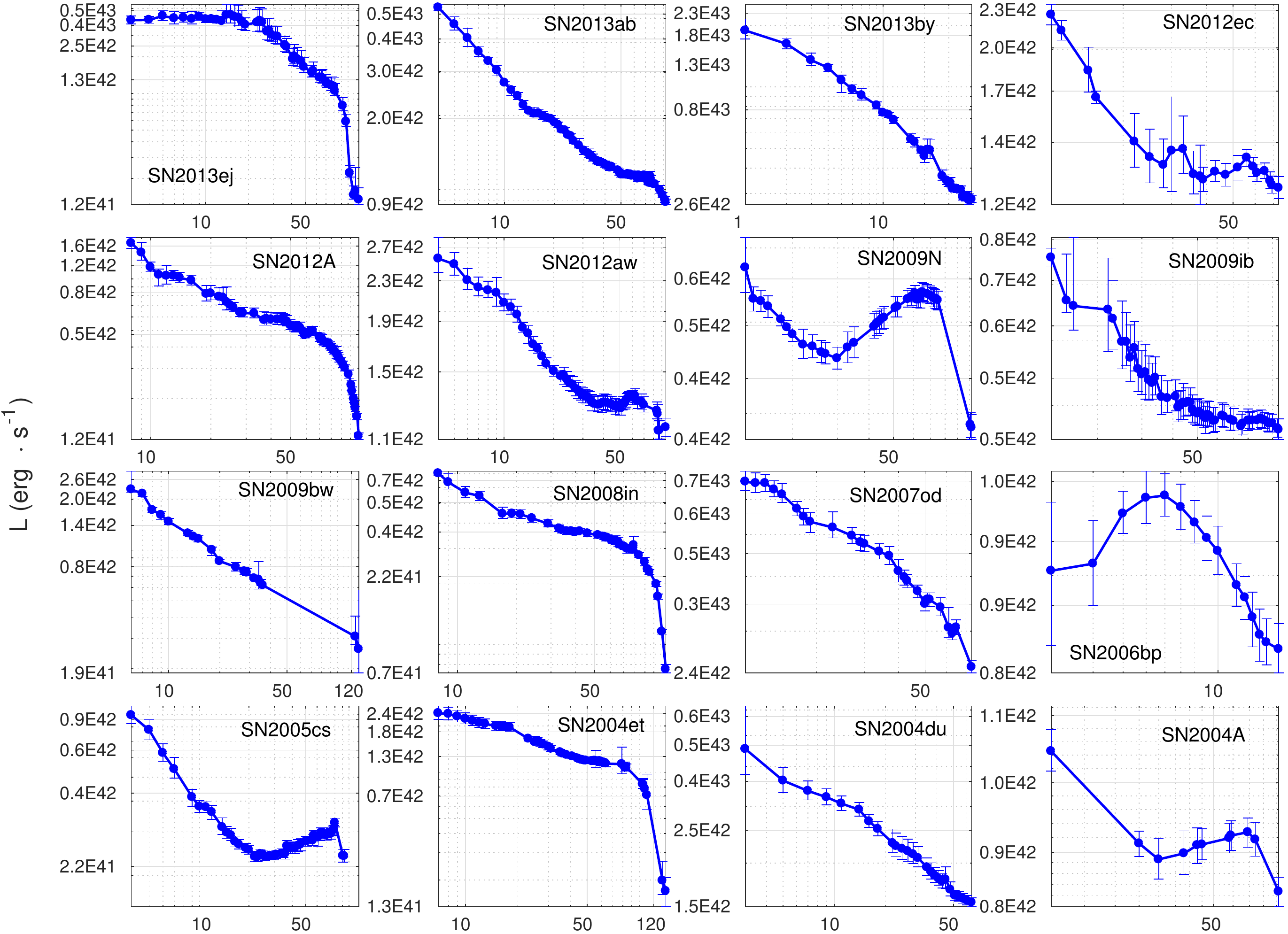} 
\includegraphics[width=0.85\textwidth]{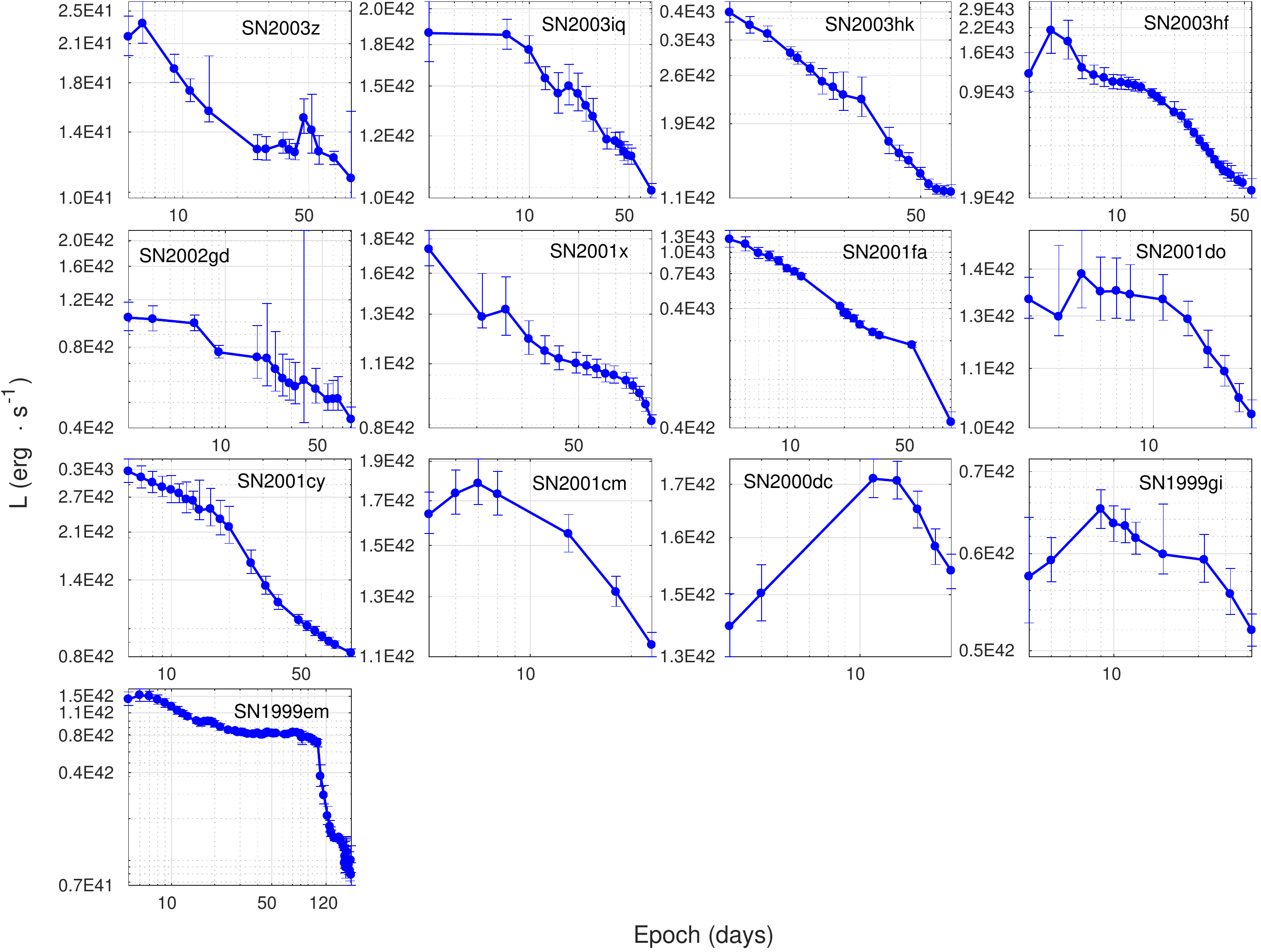} 
\caption{The bolometric luminosity curves as calculated from the black body fits for each SN in the sample.}\label{f:L}
 \end{figure*}

The bolometric luminosity for each of the SNe is computed from the fit and the curves are presented in Figure. \ref{f:L}. Similarly to the temperature, the luminosity typically decreases as a power law during early epochs. The luminosity in most of the objects relents from its fast decline and starts to decrease more moderately, where the flattening seems to coincide with the break in the temperature. There are 3 objects whose luminosity not only flattens but also starts to rise. This happens for SN2004A and SN2009N at day 30 after explosion, and for SN2005cs at day 23. The transition in luminosity happens quite sharply and occurs when the temperatures are 6000K, 5900K and 6900K for the 3 objects, respectively. The change in the evolution of the luminosity is probably also related to the recombination of the envelope. We will discuss this further in Section \ref{s:theory}. At the end of the plateau, the bolometric luminosity falls sharply.

\begin{figure*}
 \centering
\includegraphics[width=1\textwidth]{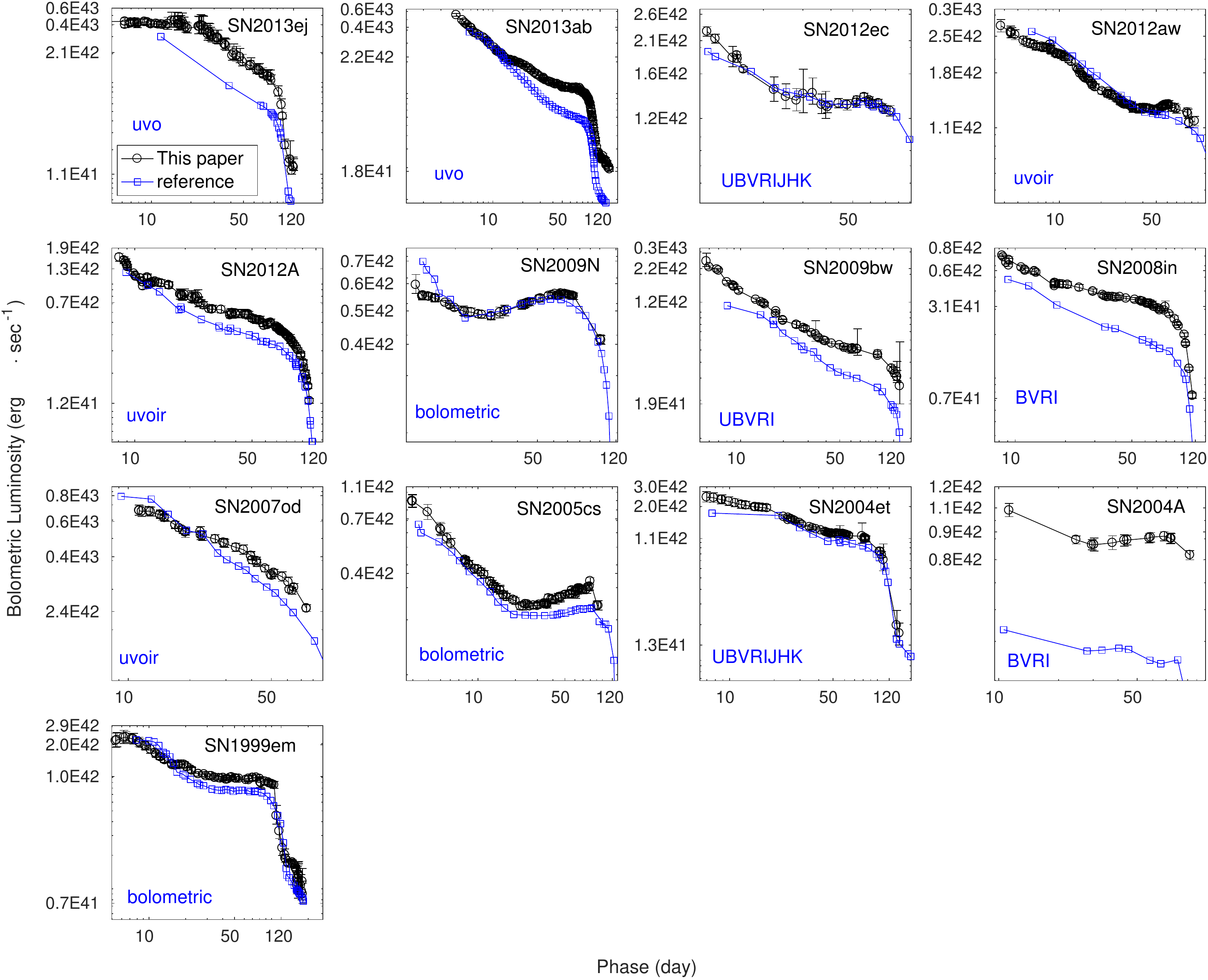} 
\caption{Comparison between the bolometric luminosity curves calculated in this paper (blue) and pseudo-bolometric curves from the literature (red), where the integrated wavelength range is specified. $'bolometric'$ light curves include bolometric corrections or are based on black body fits. Otherwise, the light curves are the integrated luminosity in the observed bands without any bolometric corrections. The luminosity of SN1999em presented here was calculated after correcting for $\rm A_{V}^{host}=0.18$, in order to compare to \citet{Bersten2009}. Discrepancies between the curves at early times are probably due to missing UV flux in the pseudo-bolometic curves. SN2012ec \citep{Barbarino2015}, SN2012aw \citep{DallOra2014}, SN2012A \citep{Tomasella2013}, SN2013ab \citep{Bose2015a}, SN2009N (flux in RJ was approximated by RJ tail, no corrections in the blue) \citep{Takats2014}, SN2009bw \citep{Inserra2012a}, SN2008in \citep{Roy2012}, SN2005cs (bolometric corrections) \citep{Pastorello2009}, SN2004et \citep{Maguire2010b}, SN2004A \citep{Hendry2006,Maguire2010b}, SN1999em \citep{Bersten2009}, SN2007od \citep{Inserra2011}, SN2013ej \citep{Bose2015b}} \label{f:L_bol_compare}
 \end{figure*}

We compare our bolometric luminosity curves to pseudo-bolometric curves from the literature by correcting for the different assumed distances to the SNe, and shifting in time to match the assumed explosion day. 
While broadly speaking there is mostly agreement between our work and previous efforts, there are still some discrepancies. The comparison is presented in Figure. \ref{f:L_bol_compare}. Pseudo-bolometric luminosities that were not computed with UV nor JHK data, as done for SN2009bw, SN2008in, and SN2004A, can be underestimated by  up to 30$\%$.

\begin{figure*}
 \centering
\includegraphics[width=0.85\textwidth]{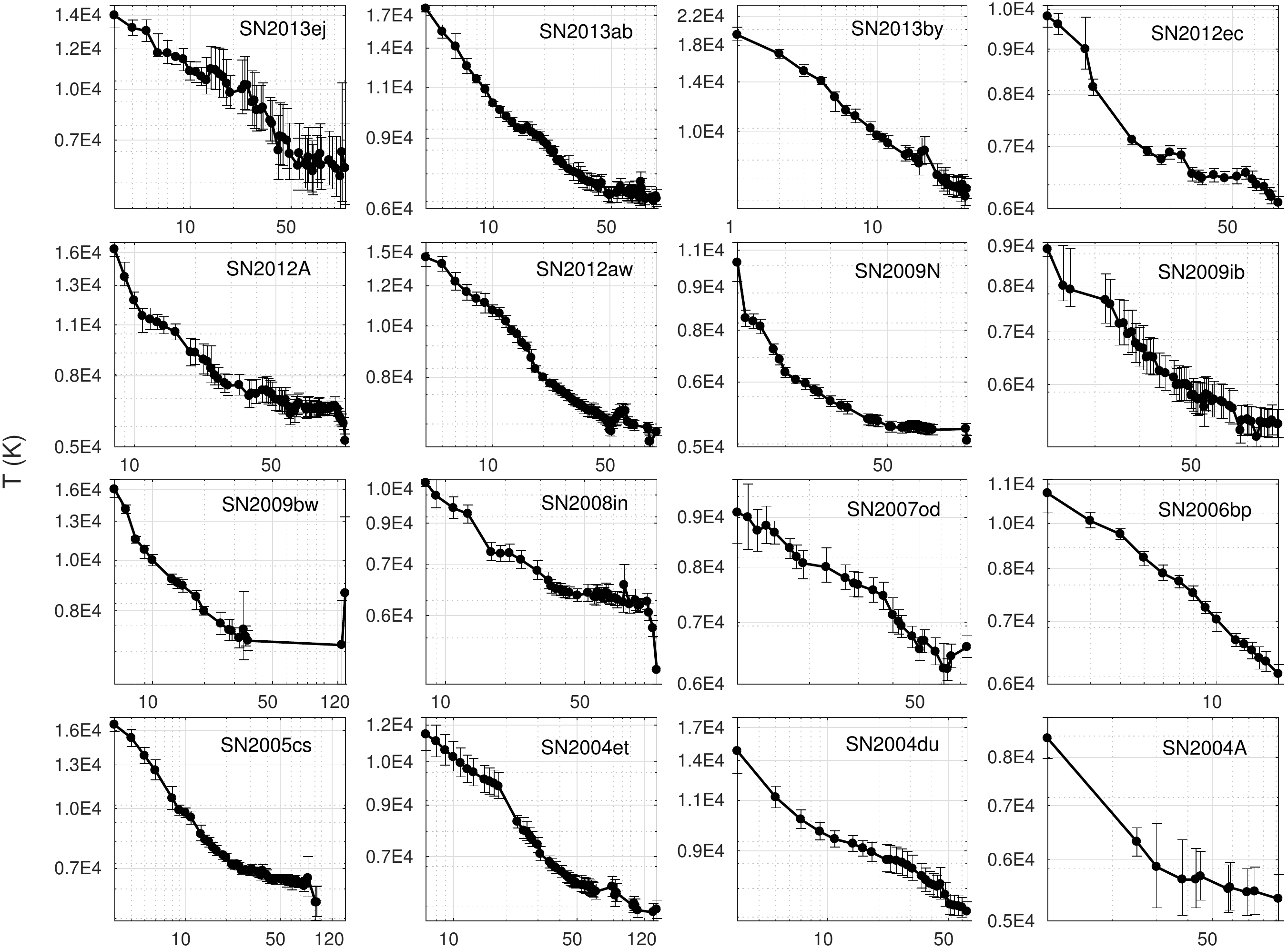} 
\includegraphics[width=0.85\textwidth]{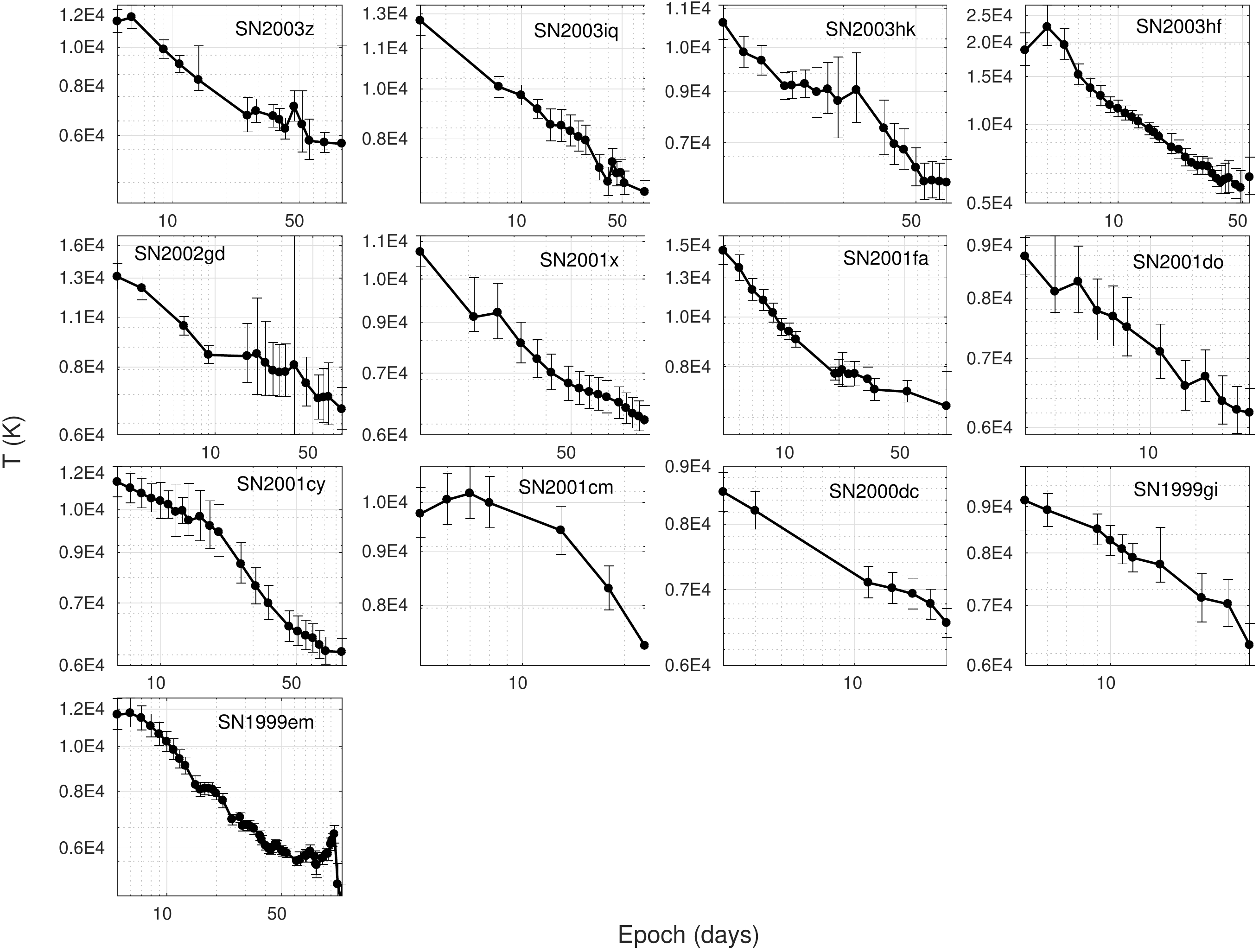} 
\caption{The temperature as a function of time for each SN in the sample.}\label{f:T}
 \end{figure*}
 
\subsection{The Effect of Extinction} 
 \label{s:extinction}
Host interstellar or circumstellar dust can introduce extinction that is not corrected for in our data (see Section \ref{s:sample}), resulting in an underestimation of the fit temperatures and luminosities. Although it is quite difficult to find a good estimation for $\rm A_{V}^{host}$, it is possible to quantify the effect a certain Av value has on the fit parameters as a function of the temperature. We repeat the fitting procedure two more times assuming E(B-V) = 0.1 and 0.05, and R$_{V}$=3.1, using the galactic extinction laws of \citet{Cardelli1989}. As most type-II SNe in or sample are expected to have $\rm E(B-V)^{host}<0.1$ (Faran et al. 2014a), this value is effectively an upper limit on the possible required corrections.

\begin{figure}
\centering
\includegraphics[width=0.45\textwidth]{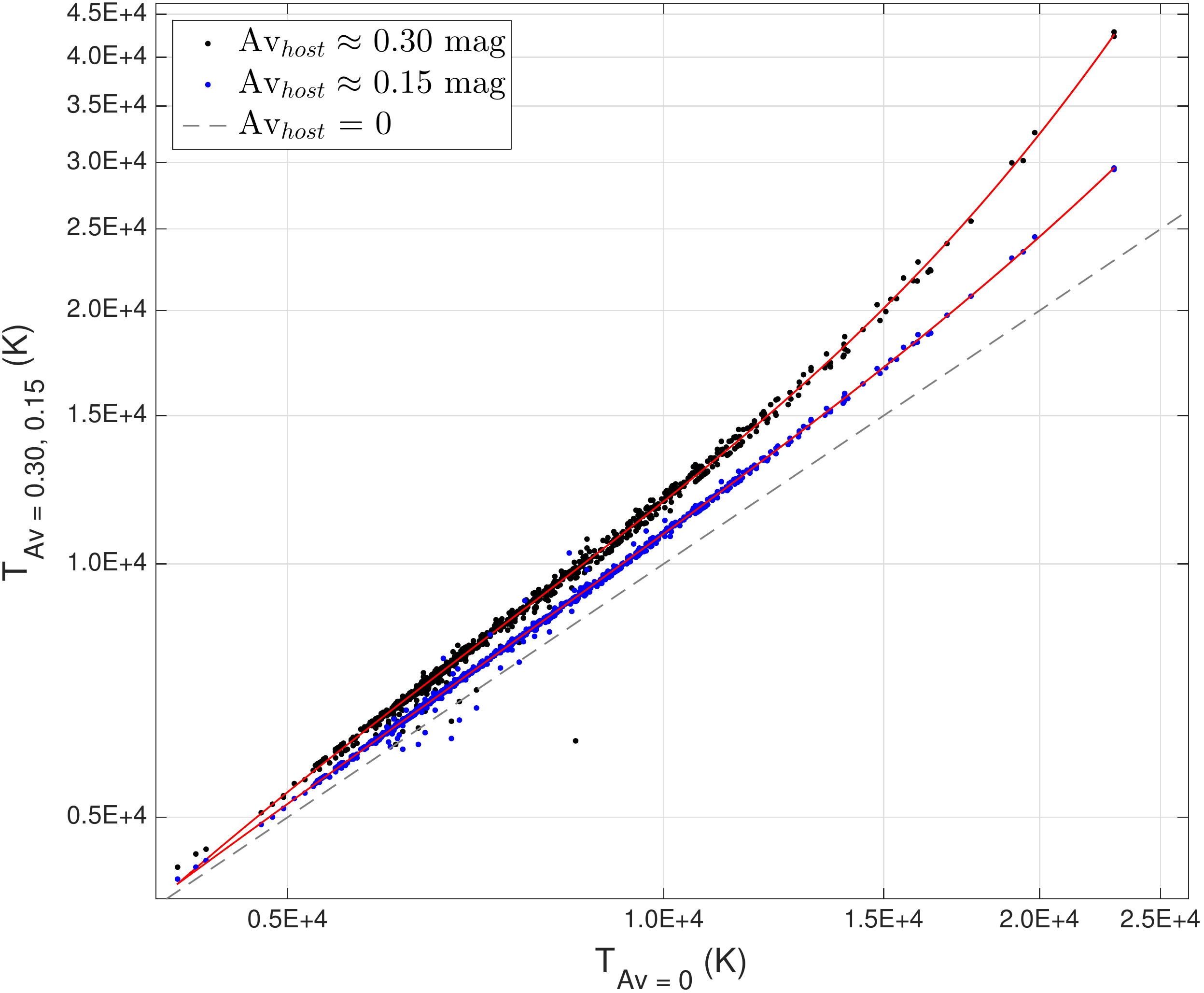} 
\caption{The best fit temperatures resuting from an extinction correction of $\rm A_{V}^{host}=0.3$ (black dots) and $\rm A_{V}^{host}=0.15$ (blue dots), as a function of $\rm T(A_{V}^{host}=0)$. The grey dashed line indicates $\rm T = T_{A_{V}^{host}=0}$. The third order polynomial fits to the data are plotted in red and can be used to translate between the uncorrected and corrected temperatures.}\label{f:T_Av}
\end{figure}

In Figure \ref{f:T_Av} we present the relation between the best fit temperatures resulting from the correction to $\rm E(B-V)^{host} = 0.1$ ($\rm A_{V}^{host}\approx 0.3$) and $\rm E(B-V)^{host} = 0.05$ ($\rm A_{V}^{host}\approx 0.1$) as a function of the uncorrected SN temperatures. The dependence of the corrected temperatures on $\rm T( A_{V}^{host}=0$) can be well described by a third order polynomial, according to the following relations:
\begin{multline}\label{eq:TAv03}
 T(A_{V}=0.3) \approx 3.1T_{3}^{3}-58T_{3}^{2}+1630T_{3}-1730
\end{multline}
and: 
\begin{multline}
T(A_{V}=0.15) \approx 0.69T_{3}^{3}-9.14\times 10^{-2}T_{3}^{2}+1150T_{3}-424,
\end{multline}
where $\rm T_{3} \equiv T(Av=0)/10^{3}$.
These relations offer a convenient way to estimate the error on a fit temperature, in the typical extinction range of $\rm A_{V}^{host}=0 - 0.3$ mag.

\begin{figure}
\centering
\includegraphics[width=0.45\textwidth]{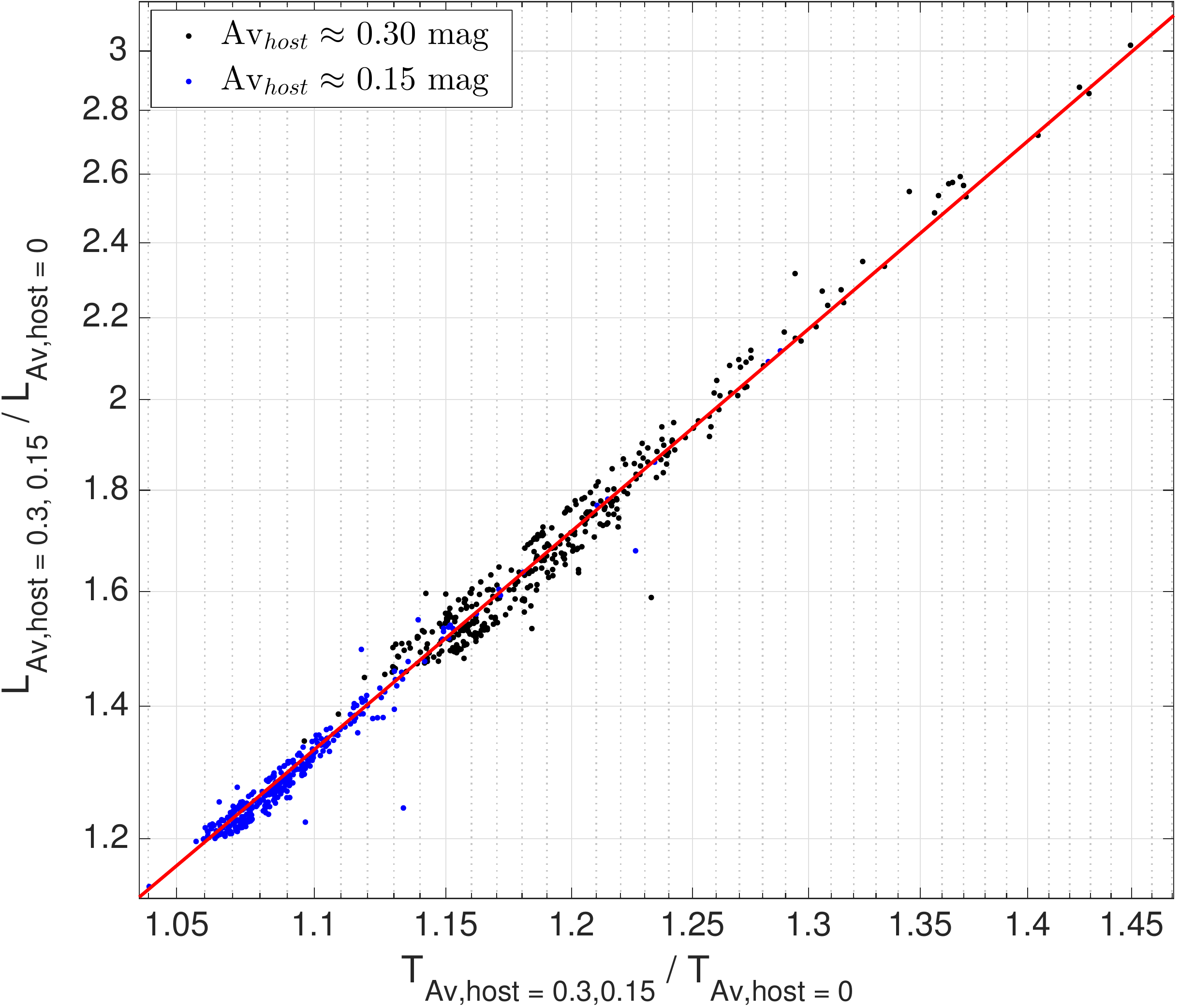} 
\caption{The ratio between the extinction corrected L$_{bol}$  ($\rm A_{V}^{host}=0.3$ in black and $\rm A_{V}^{host}= 0.15$ in blue) and the uncorrected L$_{bol}$, as a function of the similar temperature ratio for $\rm T_{A_{V}^{host}=0}>8000K$. Since extinction has a minimal effect on long wavelength observations, the Rayleigh-Jeans part of the spectrum at high temperatures is expected to be approximately fixed. Therefore, the data follow the relation: \( \frac{L_{A_{V}}}{L_{0}}  =  \Big( \frac{T_{A_{V}}}{T_{0}} \Big)^{3} \), which is presented by the red line. }\label{f:L_Av}
\end{figure}

Since the effect of extinction on the RJ is weak, we expect the luminosity to behave as \( \frac{L_{A_{V}}}{L_{0}}  =  \Big( \frac{T_{A_{V}}}{T_{0}} \Big)^{3} \) at high temperatures. We fit the data with $\rm T(A_{V}=0)>8000 K$ according to this relation for both Av=0.3 and Av=0.15 and present the data and the fit in Figure. \ref{f:L_Av}. This, together with the previous relation for the temperatures, allows also the bolometric luminosity to be corrected for extinction as the relation holds down to low temperatures of $\sim 8000K$. Below that temperature, the corrections to L are less than 10\%, which is of the order of the uncertainty.

\section{Comparison to Theory}
\label{s:theory}
\subsection{The Temperature at the Beginning of the Plateau}

The formation process of the plateau in Type-II SNe and the origin of its shape (i.e., its luminosity and temperature evolution) are not fully understood. The common wisdom states that the plateau is formed due to a recombination wave that propagates into the envelope in Lagrangian coordinates. The recombination front defines the photosphere and therefore also fixes its temperature to the temperature of hydrogen recombination in the envelope. According to this view, the plateau should start when T$\approx$7500K. However, more detailed theoretical models show that the peak in each photometric band is observed slightly before the black body peak enters the observed band. This is why redder bands peak at later time. Recombination prevents the observed temperature from falling below $\sim$6000K, which is the main reason that after the peak the luminosity in the optical and IR bands falls rather slowly, and creates what is referred to as the plateau. We therefore expect to find photospheric temperatures higher than 7500K when the plateau starts. 

We define the plateau starting time, t$_{\rm p}$, in a specific band to be the day at which the light curve changes by less than 0.02 magnitudes per day. To find t$_{\rm p}$, we fit a low order polynomial to the first 15-20 days and find the day where the derivative equals 0.02 mag/day. In order to estimate the uncertainties in t$_{\rm p}$, we use the photometric errors of the data to generate random Gaussians errors, from which we create simulated data. We run the fit 1000 times on simulated data and use the mean of the results as the value of t$_{\rm p}$ and the standard deviation as its uncertainty. The value of t$_{\rm p}$ can be sensitive to the order of the polynomial and to the time range chosen for the fit. The maximal discrepancies introduced by changing those parameters are typically not larger than one day. We therefore set a minimal error of one day on t$_{\rm p}$. 

\begin{figure}
\centering
\includegraphics[width=0.45\textwidth]{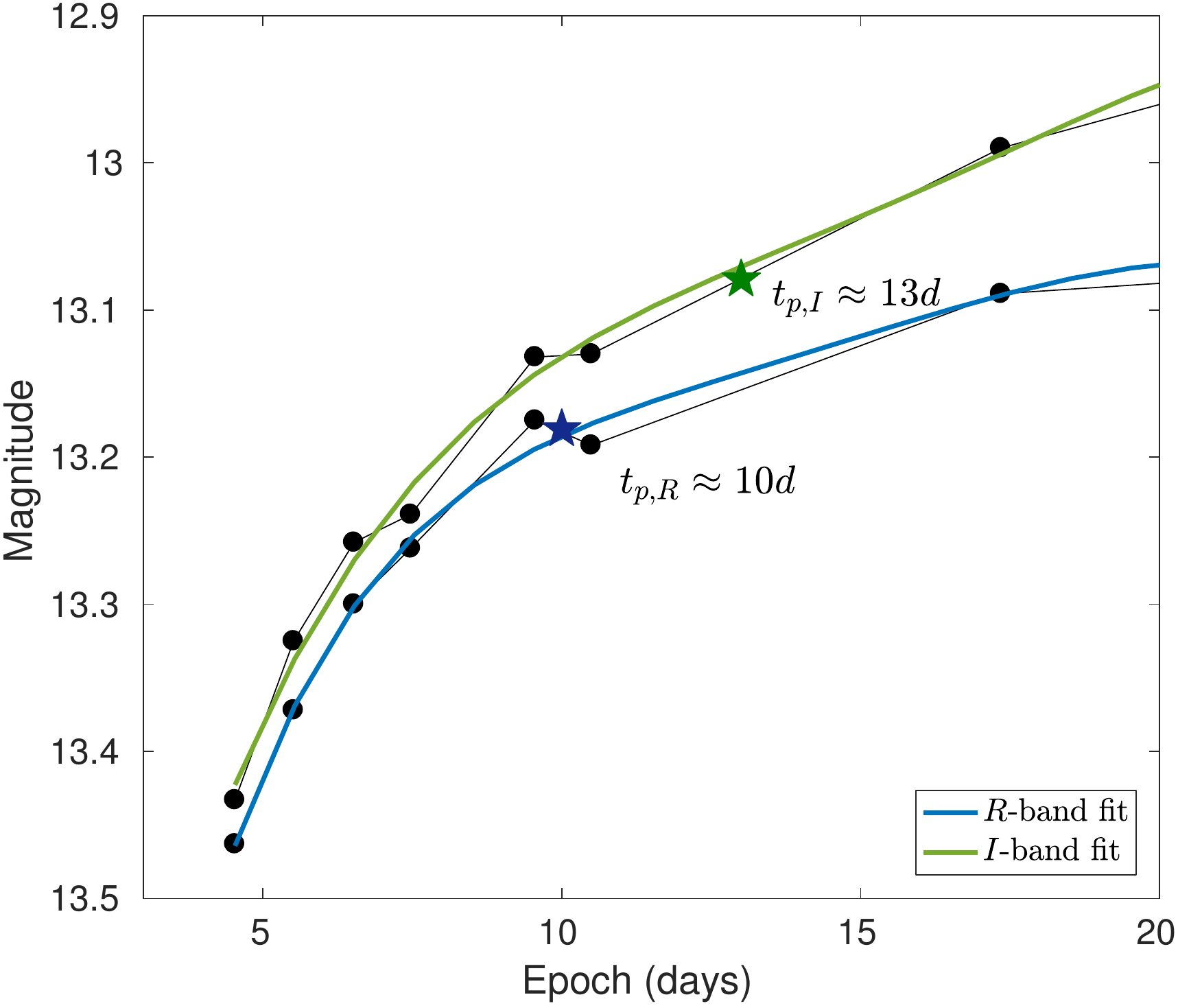} 
\caption{The locations of t$_{p,R}$ and t$_{p,I}$ of SN2012aw, as defined by the derivative of the polynomial fit to the $R$-band (blue) and the $I$-band (green). The plateau in the $I$-band appears to start slightly later than the plateau in the $R$-band.}\label{f:tp_R_vs_I}
\end{figure}

In Table \ref{t:T_plateau} we present the t$_{\rm p}$ values computed in the $R$ and $I$ bands. Some objects have only an upper limit on t$_{\rm p}$, since they were first observed already on the plateau. Nevertheless, for most of the objects it is clear that the plateau in $R$ starts slightly before the plateau in $I$, as predicted by theory. In Figure \ref{f:tp_R_vs_I} we demonstrate the different locations of the plateau in the $R$ and in the $I$ band for SN2012aw.

\begin{figure*}
\centering
\includegraphics[width=1\textwidth]{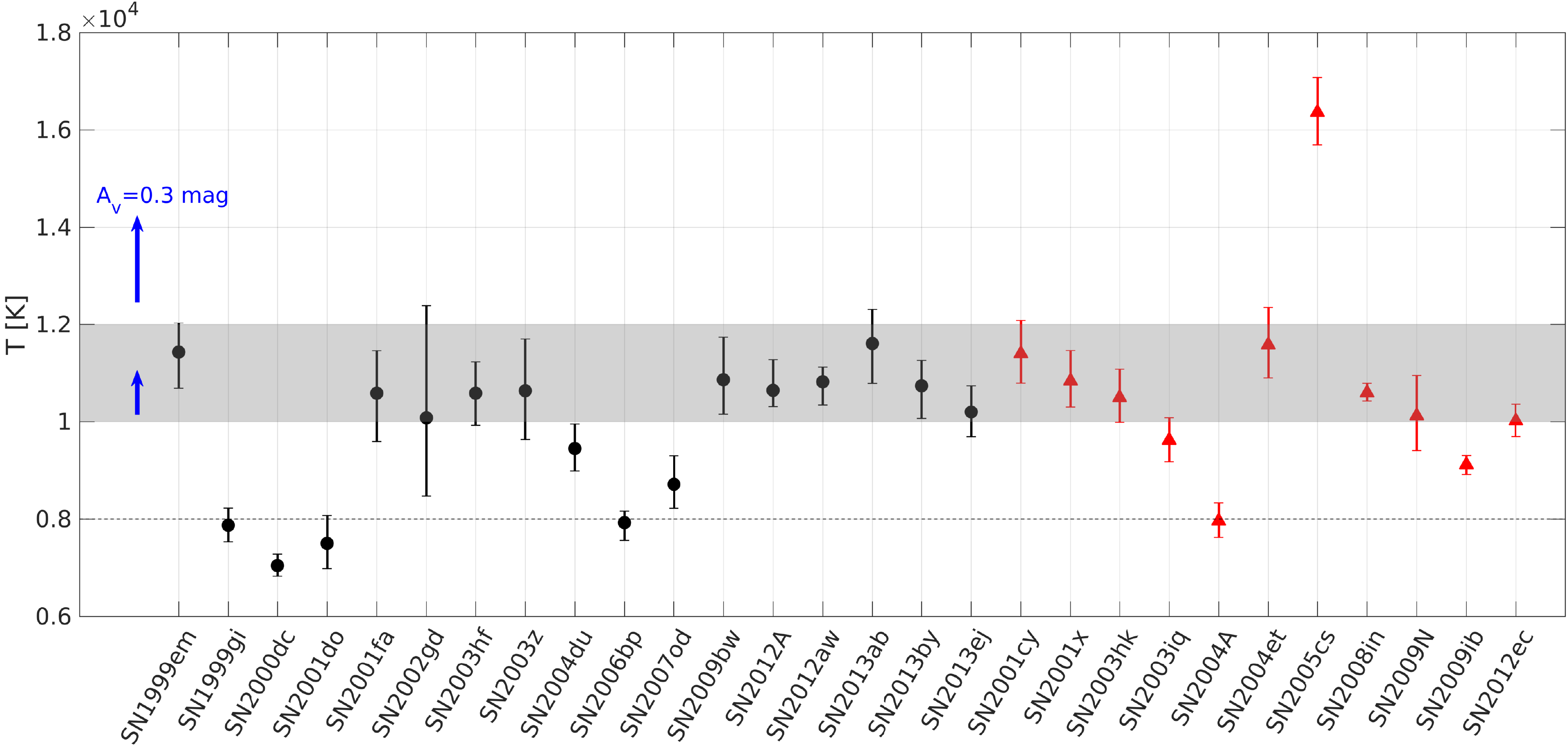} 
\caption{The temperatures at the onset of the plateau phase in the $R$ band. Lower limits are marked by red triangles. The onset of the plateau is defined as the day in which the light curve changes by less than 0.02 magnitudes per day. Most of the temperatures lie above 8000K, which reinforced our claim that the flattening of the light curve is not caused by recombination. The blue arrows show the effect of extinction on T$_{p}$ at T$\approx$11000K and T$\approx$8000K for A$_{V}$=0.3. The plateau temperatures agree with radiative transfer results calculated by \citet{Shussman2016a} for  a set of 124 RSGs, represented by the shaded area.}\label{f:Tp}
\end{figure*}

The temperatures associated with t$_{\rm p}$ in $R$ and $I$ are computed by interpolating the temperature curves to t$_{\rm p}$. We plot the temperatures at $\rm t_{p}$, i.e. T$_{\rm p} = T(t=t_{p})$ in the $R$-band for each SN in Figure. \ref{f:Tp}. The blue arrows indicate the effect that $\rm A_{V}^{host}=0.3$ would have on T$_{p}$ at $\rm T \approx 8000K$ and $\rm T \approx 11,000 K$, according to equation \ref{eq:TAv03}. Objects with only lower limits (i.e., first data point lies already on the plateau) are presented by red triangles. Almost all T$_{p}$ values lie above 8000K, and many of them above 10,000K. The low luminosity SN2005cs shows an exceptionally high lower limit of $T_{\rm p} \gtrsim 16,500$K. The observed range of $T_p$ (with the exception of SN2005cs) is consistent with the theoretical light curves prediction by \citet{Shussman2016b}. For example, the predicted $R$-band $T_p$  for explosion energy of $10^{51}$ erg of progenitors with radii in the range of $400-800~R_\odot$ and ejecta masses in the range of $7-15~M_\odot$ is between about 10,000K and 12,000K.

\begin{table*}
\def\arraystretch{1.3}
\begin{tabular}{lcccc}
\hline
SN name&t$_{p}$ $R$-Band&T$_{p}$ $R$-band&t$_{p}$ $I$-Band&T$_{p}$ $I$-band\\ 
\hline
SN1999em&8.0 (1.0) &11400$^{+600}_{-800}$&8.0 (1.1)  &11400$^{+700}_{-800}$  \\ 
SN1999gi&10.0 (1.0) &7900$^{+400}_{-400}$&10.0 (1.0)  &7900$^{+400}_{-400}$  \\ 
SN2000dc&13.0 (1.0) &7000$^{+300}_{-300}$&13.0 (1.0)  &7000$^{+300}_{-300}$  \\ 
SN2001cm&&&6.0*&9600$^{+500}_{-500}$* \\ 
SN2001cy&6.0*&11400$^{+700}_{-700}$*&6.0*&11400$^{+700}_{-700}$* \\ 
SN2001do&8.0 (1.0) &7500$^{+600}_{-600}$&11.0 (1.0)  &7100$^{+600}_{-500}$  \\ 
SN2001fa&8.0 (1.0) &10600$^{+900}_{-1000}$&9.0 (1.0)  &10000$^{+1000}_{-700}$  \\ 
SN2001x&13.0*&10800$^{+700}_{-600}$*&13.0*&10800$^{+700}_{-600}$* \\ 
SN2002gd&6.0 (3.0) &10000$^{+2400}_{-1700}$&6.0 (3.0)  &10000$^{+2400}_{-1700}$  \\ 
SN2003hf&13.0 (1.0) &10600$^{+700}_{-700}$&14.0 (1.1)  &10400$^{+700}_{-700}$  \\ 
SN2003hk&15.0*&10500$^{+600}_{-600}$*&16.0*&10500$^{+600}_{-600}$* \\ 
SN2003iq&7.0*&9600$^{+500}_{-500}$*&7.0*&9600$^{+500}_{-500}$* \\ 
SN2003z&9.0 (1.5) &11000$^{+1100}_{-1100}$&12.0 (1.6)  &9300$^{+700}_{-800}$  \\ 
SN2004A&11.0*&8000$^{+400}_{-400}$*&11.0*&8000$^{+400}_{-400}$* \\ 
SN2004du&10.0 (1.0) &9500$^{+500}_{-500}$&11.0 (1.0)  &9300$^{+500}_{-500}$  \\ 
SN2004et&9.0*&11600$^{+800}_{-700}$*&9.0*&11600$^{+800}_{-700}$* \\ 
SN2005cs&3.0*&16400$^{+800}_{-700}$*&3.0*&16400$^{+800}_{-700}$* \\ 
SN2006bp&8.0 (1.0) &7900$^{+300}_{-400}$&8.0 (1.0)  &7900$^{+300}_{-400}$  \\ 
SN2007od&9.0 (1.0) &8700$^{+600}_{-600}$&9.0 (1.0)  &8700$^{+600}_{-600}$  \\ 
SN2008in&9.0*&10600$^{+200}_{-200}$*&9.0*&10600$^{+200}_{-200}$* \\ 
SN2009N&14.0*&10100$^{+900}_{-800}$*&14.0*&10100$^{+900}_{-800}$* \\ 
SN2009bw&9.0 (1.0) &10900$^{+900}_{-800}$&11.0 (1.0)  &9800$^{+600}_{-600}$  \\ 
SN2009ib&13.0*&9100$^{+200}_{-300}$*&13.0*&9100$^{+200}_{-300}$* \\ 
SN2012A&13.0 (1.7) &10600$^{+700}_{-400}$&14.0 (1.0)  &10400$^{+600}_{-500}$  \\ 
SN2012aw&10.0 (1.1) &10800$^{+300}_{-500}$&13.0 (3.1)  &10000$^{+900}_{-1200}$  \\ 
SN2012ec&15.0*&10000$^{+400}_{-400}$*&15.0 *&10000$^{+400}_{-400}$* \\ 
SN2013ab&9.0 (1.0) &11600$^{+800}_{-900}$&9.0 (1.0)  &11600$^{+800}_{-900}$  \\ 
SN2013by&8.0 (1.0) &10700$^{+600}_{-700}$&8.0 (1.0)  &10700$^{+600}_{-700}$  \\ 
SN2013ej&12.0 (1.6) &10200$^{+600}_{-600}$&11.0 (1.4)  &10400$^{+600}_{-600}$  \\ 
\hline 
\end{tabular}
\caption{The times at which the $R$- and the $I$-bands enter the plateau phase, and the corresponding temperatures at those epochs.}

\label{t:T_plateau}
\end{table*}

\subsection{Signs of Recombination in the Temperature and Luminosity Curves}
\label{s:logarithmic_deriv}

As discussed in Sections \ref{s:Temperature} and \ref{s:Luminosity}, the evolution of the temperature and the 
bolometric luminosity is characterized well by a power-law, that flattens when the temperature drops to 
$\sim$6000--7000K. We compute the early values of the logarithmic derivatives of the luminosity and temperature, $\rm \alpha_L$ and $\rm\alpha_T$, respectively, during the first 15 days after the explosion. SNe that do not have U or UV data are excluded, since the temperatures at these epochs are typically higher than 12,000K, where U-band data (or bluer) are important to constrain the fit (see Section \ref{s:fitting}). We also calculate the late logarithmic derivatives between 40 and 100 days, while the SN light curve is on the plateau. For this we choose only SNe with IR data for the reasons discussed in Section \ref{s:fitting}. The results are summarized in Table \ref{t:alpha_TL}. The best fit values for the early power law are highly sensitive to the exact value of the zero point in time. Since we make conservative explosion day estimates (see Section. \ref{s:sample}), some of the uncertainties on the explosion day are as large as 5-10 days, and introduce non-negligible uncertainties to the values of the power law. The uncertainty values introduced from the fit itself and from the uncertainty on the explosion day are presented separately in Table \ref{t:alpha_TL}. The values in the parentheses are the errors produced by the fit, and the upper and lower values are the differences from the $\rm \alpha$ values that we get using the lower and upper boundaries of the explosion day estimate, respectively. In cases where the explosion day uncertainty is large (as in SN2013by, SN2012ec, SN2009jb and SN2008in) the upper and lower uncertainties are quite large. However, the explosion day uncertainty naturally has very little effect on the late values of $\rm \alpha$.

\begin{figure}
\centering
\includegraphics[width=0.5\textwidth]{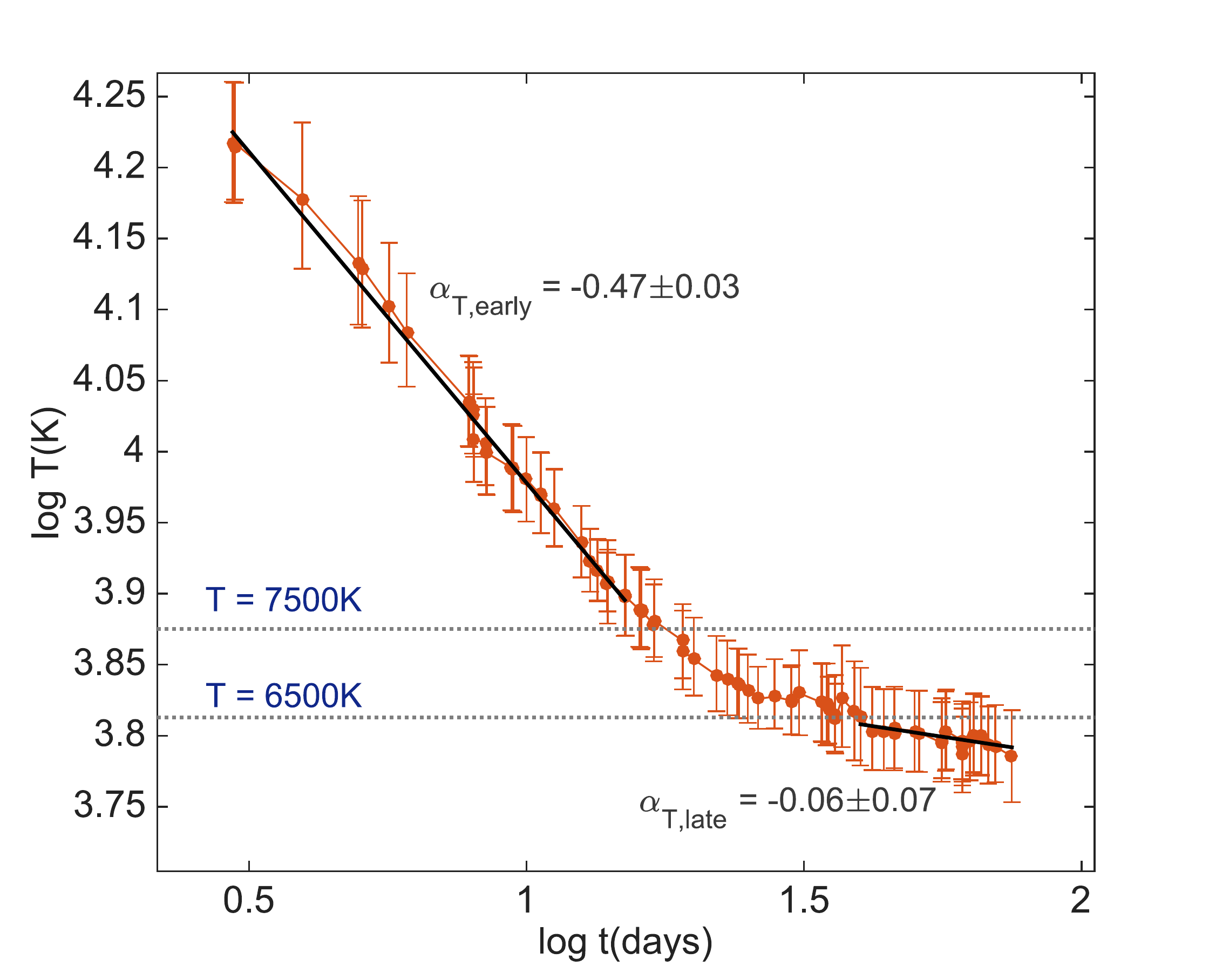} 
\caption{The temperature curve of SN2005cs on logarithmic scale. The temperature behaves as a broken power law, where in early phases it declines rapidly with a power of $\rm -0.47 \pm 0.03$, and at late phases the power law is $\rm -0.06 \pm 0.07$. This transition in the power law happens within the temperature range 6500--7500K, and we interpret it as the time when the color shell reaches the recombination temperature. }\label{f:alpha_T_05cs}
\end{figure}

\begin{figure*}
\centering
\includegraphics[width=1\textwidth]{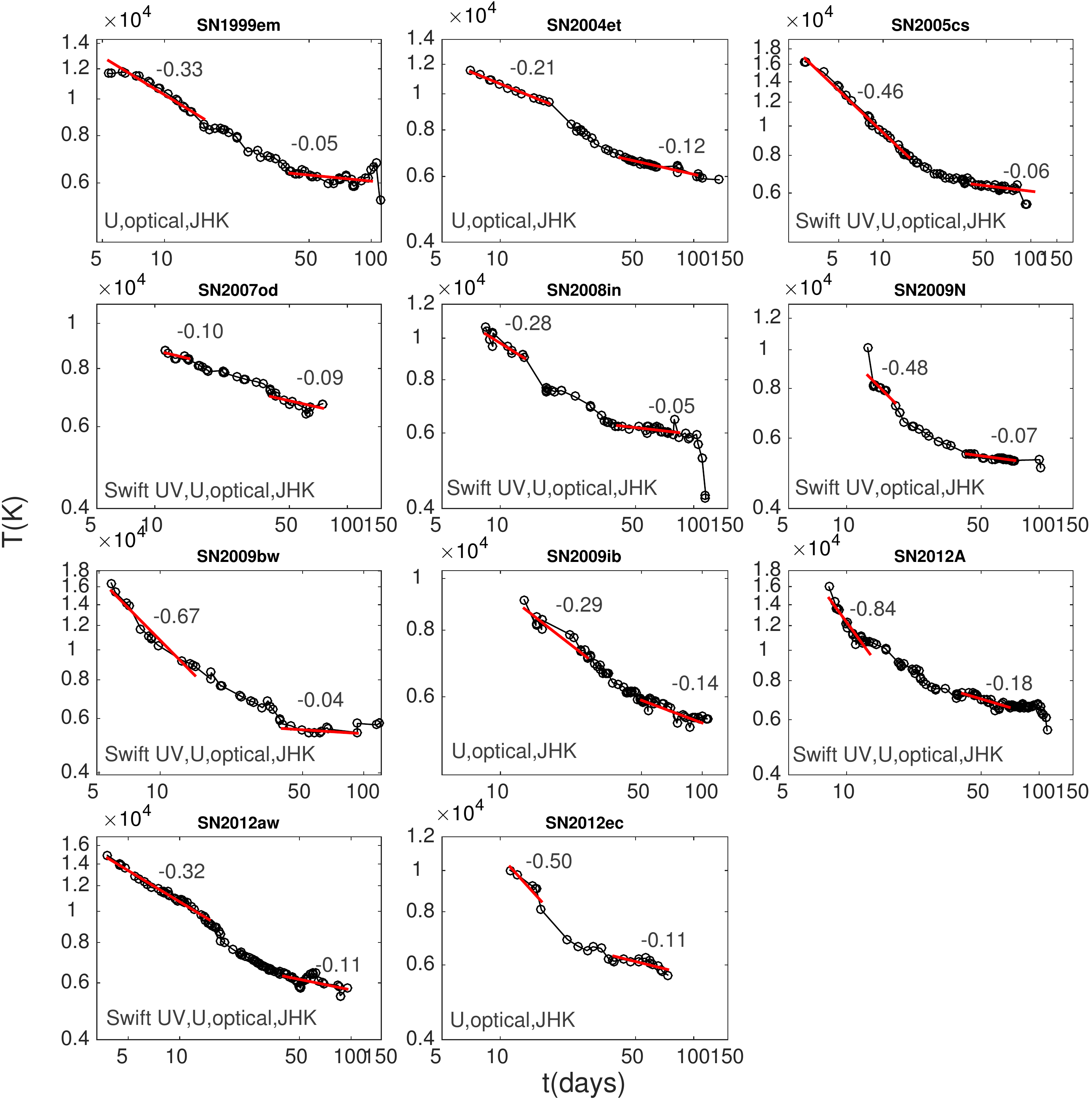} 
\caption{Temperature curves of objects that have both UV and JHK data. The numbers indicate the values of the best-fit early and late logarithmic derivative computed during the first 15 days, and during days 40-100 after the explosion, respectively. A clear flattening of the temperature is observed as the SN approcahes the temperaure of hydrogen recombination.}\label{f:alpha_T}
\end{figure*}

An example of the fit for $\alpha_T$ is shown in Figure. \ref{f:alpha_T_05cs}. It depicts the temperature curve of SN2005cs, on a logarithmic sacale. The best-fit logarithmic derivative computed during the first 15 days is $\rm\alpha_T=-0.47\pm0.03$, and during days 40-100 is $\rm \alpha_T=-0.06\pm0.07$. There is a clear flattening of the temperature curve between t=19d and t=35d, when the temperature is between 6500K and 7500K. In Figure \ref{f:alpha_T} we present the temperature curves and the power law fits for all the objects that have both UV and IR data. From the values of the logarithmic derivatives (table \ref{t:alpha_TL}) it is clear that at some point the temperature evolution flattens. At early time most values are in the range $\rm \alpha_{T,early} \sim$-0.6 -- -0.2 while at late time all best fit values are in the range $\rm \alpha_{T,late} \sim$-0.15 -- 0. The weighted mean values of the logarithmic derivatives are $\rm \bar{\alpha}_{T,early} = -0.38 \pm 0.01$ and $\rm \bar{\alpha}_{T,late} = -0.08 \pm 0.02$. Although it is not possible to point out the exact temperature of the transition, one can see that the range of temperatures between the two power law regimes is $\sim$6000-7000K, which is the temperature range expected from hydrogen recombination in type-II SN envelopes. 

We also calculate the early and late logarithmic derivatives of L$_{\rm bol}$. Similar to the temperature, the bolometric luminosity curves generally have a higher logarithmic derivative in the early phases. Most of the values of $\alpha_{\rm L,early}$ are  between -0.2 and -0.8, while most values of $\alpha_{\rm L,late}$ are  between -0.6 and 0.2, with weighted mean values of $\rm \bar{\alpha}_{L,early} = -0.46 \pm 0.01$ and $\rm \bar{\alpha}_{L,late} = -0.22 \pm 0.03$ . When including the effect of extinction, we find that the values of $\rm \alpha_T,early$ and $\rm \alpha_L,early$ increase by $\sim 0.1$ and $\sim 0.2$ respectively, with an extinction value of $\rm E(B-V)=0.1 mag$.
This result is also consistent with the expectation from recombination which is expected to cause a flatter, or even rising, light curves once the recombination front reaches facilitate the release of radiation from inner regions.

\begin{figure}
\centering
\includegraphics[width=0.5\textwidth]{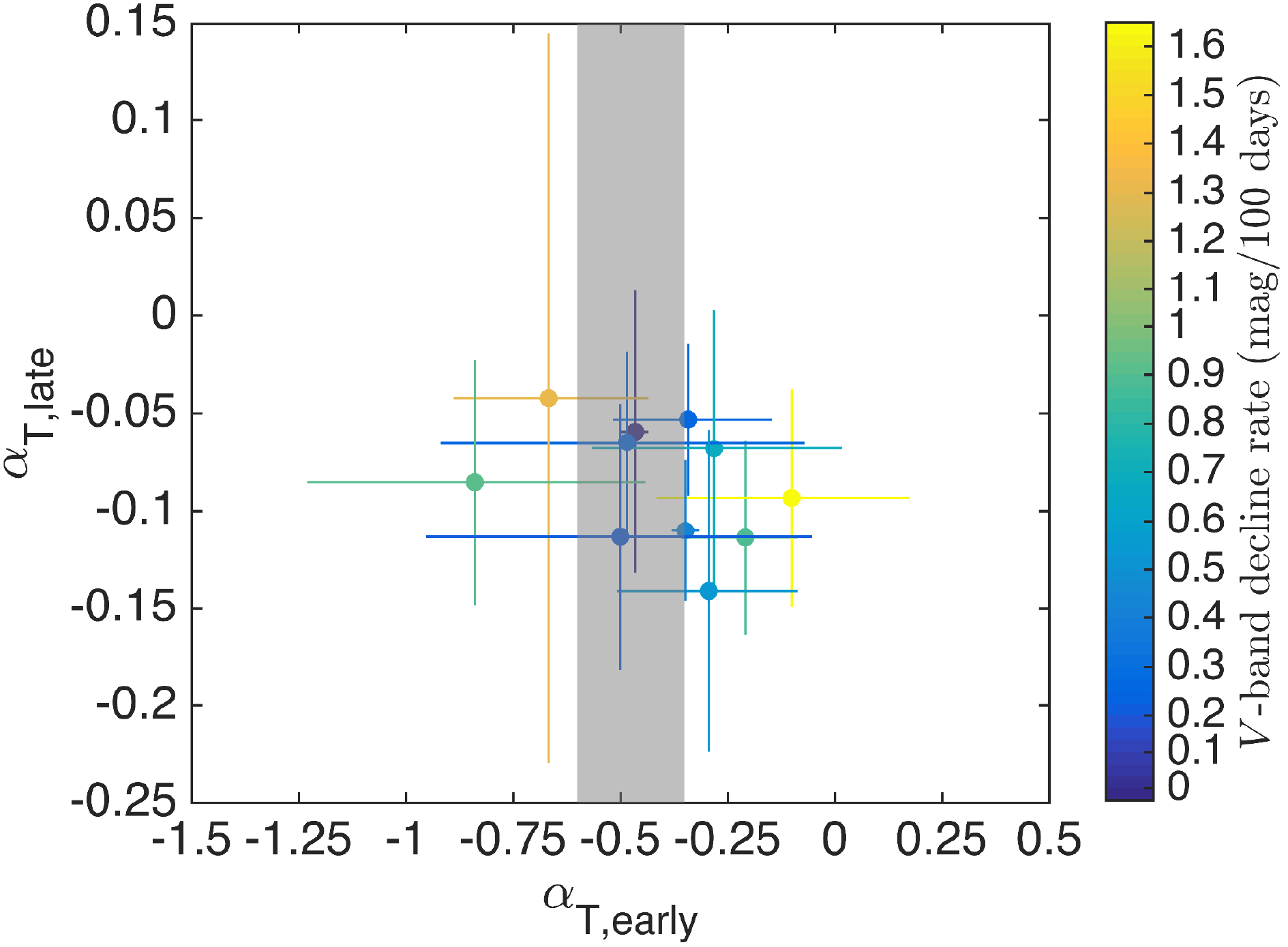} 
\includegraphics[width=0.5\textwidth]{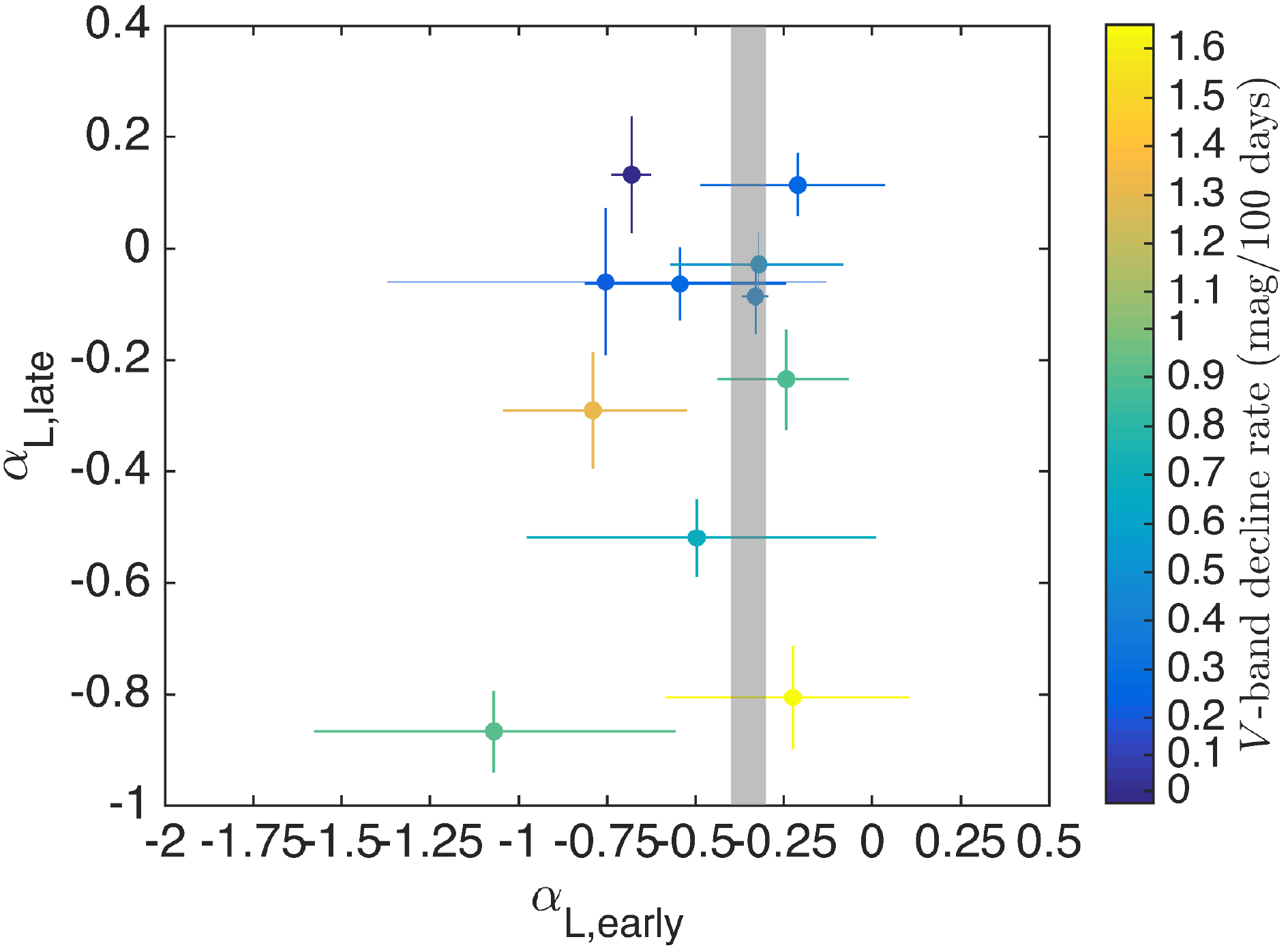} 
\caption{Top: the logarithmic derivative of the temperature at late times (40-100 days after the explosion) vs. the logarithmic derivative of the temperature at early times (up to 15 days after the explosion). The colors represent the decline rate of the $V$-band light curve. The values of the different objects agree with each other within the error-bars, and there is no apparent correlation with the light curve decline rate. Bottom: the same as the top figure for the bolometric luminosity. In this case, the early values of the logarithmic derivatives agree within the errors, but the late values show a wider spread. SNe whose luminosity declines faster at late phases (i.e., during the plateau phase) have faster declining light curves.}\label{f:alpha_TL_early_late}
\end{figure}

An interesting question is whether there are correlations between the early and late evolution, or between temperature and luminosity evolution.  In Figure \ref{f:alpha_TL_early_late} we plot $\alpha_{\rm T,late}$ vs. $\alpha_{\rm T,early}$ (upper panel) and $\alpha_{\rm L,late}$ vs. $\alpha_{\rm L,early}$ (lower panel). The color coding refers to the decline rate of the $V$-band light curve per 100 days, calculated by linearly fitting the magnitude decline rate between day 25 and 75. The figures show no clear correlations between early and late evolution or between the late decline rate and the temperature evolution (early or late). However, there is a linear correlation between $\rm \alpha_T$ and $\rm \alpha_L$.

\citet{Shussman2016b} provides theoretical predictions of $\alpha_{\rm T,early}$ and $\alpha_{\rm L,early}$ based on numerical simulations of SN explosions of a  large set of RSG progenitors. They find that before recombination $\alpha_{\rm T}$ is not strictly constant, and that it makes a transition from about -0.35 to 
-0.6. The time of steepening in $\alpha_{\rm T}$ depends on the progenitor radius and ejecta velocity and for typical parameters it ranges between a day and two weeks. Since the data we have is not detailed enough to see the transition between the two power-laws, but only a single average power-law index, the analytic prediction is $\alpha_{\rm T,early} \approx$ -0.35 -- -0.6. This range is marked in figure \ref{f:alpha_TL_early_late} and it is broadly consistent with the observed values listed in table \ref{t:T_plateau}. The theoretical model for the luminosity evolution predicts  $\alpha_{\rm L,early} \approx -0.35$. This value depends slightly on the progenitor radius (up to about $\pm 0.05$) and more strongly on the progenitor structure (i.e., density profile). comparison of this prediction to the values listed in table \ref{t:T_plateau} shows that they are consistent for most SNe but not for all. Moreover, the value of $\alpha_{\rm L,early}$ is inconsistent with being similar to all SNe. This is probably due to different density profiles of the progenitors.

\begin{table*}
\def\arraystretch{1.3}
\centering

\begin{tabular}{lcccc}
\hline
SN name & $\alpha_{T,early}$ & $\alpha_{T,late}$ & $\alpha_{L,early}$ & $\alpha_{L,late}$ \\
\hline
SN1999em&-0.34 (0.06)$^{+0.18} _{-0.16} $&-0.05 (0.04)$^{+0.00} _{-0.00} $&-0.54 (0.09)$^{+0.28} _{-0.25} $&-0.06 (0.06)$^{+0.01} _{-0.01} $\\ 
 SN2004et&-0.21 (0.12)$^{+0.01} _{-0.01} $&-0.11 (0.05)$^{+0.00} _{-0.00} $&-0.24 (0.17)$^{+0.01} _{-0.01} $&-0.24 (0.09)$^{+0.00} _{-0.00} $\\ 
 SN2005cs&-0.47 (0.03)$^{+0.00} _{-0.00} $&-0.06 (0.07)$^{+0.00} _{-0.00} $&-0.68 (0.05)$^{+0.00} _{-0.00} $&0.13 (0.10)$^{+0.00} _{-0.00} $\\ 
 SN2006bp&-0.27 (0.01)$^{+0.03} _{-0.03} $&&-0.04 (0.01)$^{+0.01} _{-0.01} $&\\ 
 SN2007od&-0.10 (0.27)$^{+0.06} _{-0.06} $&-0.09 (0.06)$^{+0.01} _{-0.01} $&-0.22 (0.29)$^{+0.14} _{-0.14} $&-0.81 (0.09)$^{+0.12} _{-0.12} $\\ 
 SN2008in&-0.28 (0.10)$^{+0.28} _{-0.26} $&-0.07 (0.07)$^{+0.01} _{-0.01} $&-0.50 (0.12)$^{+0.49} _{-0.46} $&-0.52 (0.07)$^{+0.09} _{-0.09} $\\ 
 SN2009N&-0.48 (0.22)$^{+0.35} _{-0.34} $&-0.07 (0.05)$^{+0.01} _{-0.01} $&-0.21 (0.19)$^{+0.15} _{-0.15} $&0.11 (0.05)$^{+0.02} _{-0.02} $\\ 
 SN2009bw&-0.67 (0.05)$^{+0.23} _{-0.21} $&-0.04 (0.19)$^{+0.00} _{-0.00} $&-0.79 (0.07)$^{+0.25} _{-0.24} $&-0.29 (0.10)$^{+0.01} _{-0.01} $\\ 
 SN2009ib&-0.29 (0.12)$^{+0.16} _{-0.15} $&-0.14 (0.08)$^{+0.02} _{-0.02} $&-0.32 (0.15)$^{+0.18} _{-0.17} $&-0.03 (0.05)$^{+0.00} _{-0.00} $\\ 
 SN2012A&-0.84 (0.11)$^{+0.38} _{-0.37} $&-0.09 (0.06)$^{+0.01} _{-0.01} $&-1.07 (0.18)$^{+0.48} _{-0.47} $&-0.87 (0.07)$^{+0.06} _{-0.06} $\\ 
 SN2012aw&-0.35 (0.02)$^{+0.02} _{-0.02} $&-0.11 (0.04)$^{+0.00} _{-0.00} $&-0.33 (0.03)$^{+0.02} _{-0.02} $&-0.09 (0.06)$^{+0.00} _{-0.00} $\\ 
 SN2012ec&-0.50 (0.21)$^{+0.39} _{-0.38} $&-0.11 (0.07)$^{+0.02} _{-0.02} $&-0.75 (0.22)$^{+0.58} _{-0.56} $&-0.06 (0.13)$^{+0.01} _{-0.01} $\\ 
 SN2013ab&-0.56 (0.02)$^{+0.07} _{-0.07} $&&-0.80 (0.03)$^{+0.10} _{-0.10} $&\\ 
 SN2013by&-0.30 (0.01)$^{+0.38} _{-0.84} $&&-0.42 (0.01)$^{+0.49} _{-1.03} $&\\ 
 SN2013ej&-0.21 (0.03)$^{+0.04} _{-0.03} $&&-0.00 (0.05)$^{+0.00} _{-0.00} $&\\ 
 \hline

\end{tabular}
\caption{The early and late logarithmic derivatives of the temperature and the bolometric luminosity. The values of $\rm \alpha_{T,early}$ and $\rm \alpha_{L,early}$ increase by $\approx 0.1$  and $\approx 0.2$ respectively, with an extinction value of $\rm E(B-V)=0.1$ mag. }
\label{t:alpha_TL}
\end{table*}

\subsection{Deviation from Black Body}
\label{s:deviation_from_BB}

\begin{figure*}
\centering
\includegraphics[width=1\textwidth]{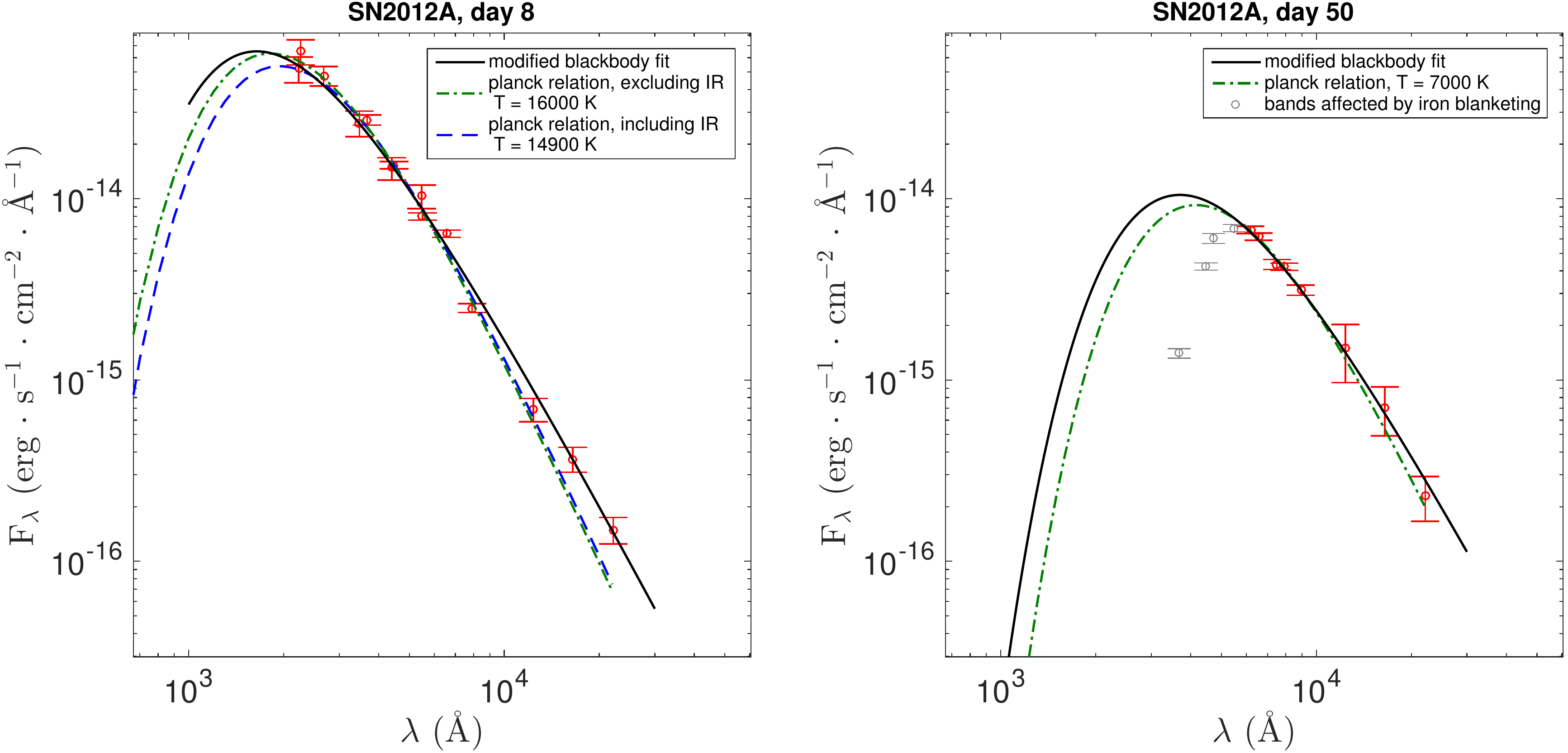} 
\caption{left:The SED of SN2012A at 8 days after the explosion. The standard Planck formula is only able to fit the peak of the distribution, but fails to fit the RJ tail. A modified black body model from \citet{Shussman2016b} is shown to be compatible throughout the whole wavelength range. right: An SED of SN2012A at 50 days past explosion. At this stage, after the onset of recombination and at low temperatures, both the modified black body and the standard model are able to describe the data. }\label{f:BB_deviation}
\end{figure*}

At high temperatures, we find that a black body is not able to describe the whole observed spectrum not only at short wavelengths, where line blanketing is important, but also at long wavelengths on the RJ tail. In the left panel of Figure. \ref{f:BB_deviation} we show the SED of SN2012A on day 8, where the temperature is $\sim$15,000K. The dashed blue line is the best fit to all of the data points, which clearly fails to fit the JHK observations. This effect is observed in all SNe that have early JHK data, which always seem to be brighter than the RJ tail at the temperature corresponding to the optical and UV flux. 

Deviation of the RJ tail from a black body spectrum of early type II emission was seen in numerical simulations \citep[e.g.,][]{Tominaga2011} and was modeled recently analytically by \citet{Shussman2016b}. According to the model the reason for the deviation is that the flux at different wavelengths is determined at different locations in the outflow at different gas temperatures. \citet{Shussman2016b} find that on the RJ tail the modified spectrum can be approximated as $F_\nu \propto \nu^{1.4}$ and they also provide an approximation for the entire spectrum. The black solid line in the left panel of Figure. \ref{f:BB_deviation} is the best fit of \cite{Shussman2016b} model to the data of SN2012A. The model seems to fit the data very well throughout the UV and the IR and follows the JHK flux where it departs from a standard black body.

At lower temperatures both the standard black body model and the modified black body models are able to describe the JHK observations, as seen in the right panel of Figure. \ref{f:BB_deviation}. The reason is that these bands are closer to the peak of the spectrum, where the models are essentially equivalent. Also, at lower temperatures, where recombination starts affecting the spectrum the analytic model is no longer applicable.

\section{Summary}
\label{s:Summary}
We calculated the temperaures and bolometric luminosities of 29 type-II SNe, by fitting black body models to their SEDs. We use the results to study the properties at the beginning of the plateau, to look for the signature of hydrogen recombination and to compare the observation before recombination becomes important to theoretical models. Our main findings are listed below.

\begin{itemize}
  \item The temperature at the onset of the plateau phase in the $R$-band is above 8000K for all SNe in our sample, and exceeds 10,000K in many of them. This temperature changes as a function of the observed band, and is determined by the temperature at which the peak of the black body spectrum roughly coincides with the center of the filter transmission curve. This result is consistent with recent theoretical models and is different than the common statment that the plateau phase starts once hydrogen recombinations becomes important. The temperatures we find agree with the predicted values for typical RSG progenitors of type-II SNe \citep{Shussman2016b}.
  \item We find that the temperature evolves with time as a power law, which flattens at $\rm \sim 6000-8000K$. We observe a similar evolution in the bolometric luminosity, where the logarithmic derivative at early phases is higher than that at late phases. The flattening is most likely a result of the recombination wave that exposes the inner layers. The values of the logarithmic derivatives for T and $\rm L_{bol}$ at early phases agree with predictions from simulations and analytic works \citep{Shussman2016b,NakarSari2010}.
  \item SN spectra deviate from a standard black body, both at low temperatures and short wavelengths due to line blanketing, and also at high temperatures and long wavelengths. We show that the SNe in our sample follow the analytic result from \citep{Shussman2016b}, that the flux on the RJ tail follows $F_\nu \propto \nu^{1.4}$.
\end{itemize}

\section*{Acknowledgements}

This research was supported by the I-Core center of excellence of the CHE-ISF. TF and EN were partially supported by an ERC starting grant (GRB/SN), an ISF grant (1277/13) and an ISA grant. 


\clearpage

\bibliographystyle{mnras}
\bibliography{Faran2017} 

\begin{thebibliography}{}
\makeatletter
\relax
\def\mn@urlcharsother{\let\do\@makeother \do\$\do\&\do\#\do\^\do\_\do\%\do\~}
\def\mn@doi{\begingroup\mn@urlcharsother \@ifnextchar [ {\mn@doi@}
  {\mn@doi@[]}}
\def\mn@doi@[#1]#2{\def\@tempa{#1}\ifx\@tempa\@empty \href
  {http://dx.doi.org/#2} {doi:#2}\else \href {http://dx.doi.org/#2} {#1}\fi
  \endgroup}
\def\mn@eprint#1#2{\mn@eprint@#1:#2::\@nil}
\def\mn@eprint@arXiv#1{\href {http://arxiv.org/abs/#1} {{\tt arXiv:#1}}}
\def\mn@eprint@dblp#1{\href {http://dblp.uni-trier.de/rec/bibtex/#1.xml}
  {dblp:#1}}
\def\mn@eprint@#1:#2:#3:#4\@nil{\def\@tempa {#1}\def\@tempb {#2}\def\@tempc
  {#3}\ifx \@tempc \@empty \let \@tempc \@tempb \let \@tempb \@tempa \fi \ifx
  \@tempb \@empty \def\@tempb {arXiv}\fi \@ifundefined
  {mn@eprint@\@tempb}{\@tempb:\@tempc}{\expandafter \expandafter \csname
  mn@eprint@\@tempb\endcsname \expandafter{\@tempc}}}

\bibitem[\protect\citeauthoryear{{Anderson} et~al.,}{{Anderson}
  et~al.}{2014}]{Anderson2014}
{Anderson} J.~P.,  et~al., 2014, \mn@doi [\apj] {10.1088/0004-637X/786/1/67},
  \href {http://adsabs.harvard.edu/abs/2014ApJ...786...67A} {786, 67}

\bibitem[\protect\citeauthoryear{{Arcavi} et~al.,}{{Arcavi}
  et~al.}{2012}]{Arcavi2012}
{Arcavi} I.,  et~al., 2012, \mn@doi [\apjl] {10.1088/2041-8205/756/2/L30},
  \href {http://adsabs.harvard.edu/abs/2012ApJ...756L..30A} {756, L30}

\bibitem[\protect\citeauthoryear{{Arnett}}{{Arnett}}{1980}]{Arnett1980}
{Arnett} W.~D.,  1980, \mn@doi [\apj] {10.1086/157898}, \href
  {http://adsabs.harvard.edu/abs/1980ApJ...237..541A} {237, 541}

\bibitem[\protect\citeauthoryear{{Barbarino} et~al.,}{{Barbarino}
  et~al.}{2015}]{Barbarino2015}
{Barbarino} C.,  et~al., 2015, \mn@doi [\mnras] {10.1093/mnras/stv106}, \href
  {http://adsabs.harvard.edu/abs/2015MNRAS.448.2312B} {448, 2312}

\bibitem[\protect\citeauthoryear{{Bersten} \& {Hamuy}}{{Bersten} \&
  {Hamuy}}{2009}]{Bersten2009}
{Bersten} M.~C.,  {Hamuy} M.,  2009, \mn@doi [\apj]
  {10.1088/0004-637X/701/1/200}, \href
  {http://adsabs.harvard.edu/abs/2009ApJ...701..200B} {701, 200}

\bibitem[\protect\citeauthoryear{{Bose} et~al.,}{{Bose}
  et~al.}{2013}]{Bose2013}
{Bose} S.,  et~al., 2013, \mn@doi [\mnras] {10.1093/mnras/stt864}, \href
  {http://adsabs.harvard.edu/abs/2013MNRAS.433.1871B} {433, 1871}

\bibitem[\protect\citeauthoryear{{Bose} et~al.,}{{Bose}
  et~al.}{2015a}]{Bose2015a}
{Bose} S.,  et~al., 2015a, \mn@doi [\mnras] {10.1093/mnras/stv759}, \href
  {http://adsabs.harvard.edu/abs/2015MNRAS.450.2373B} {450, 2373}

\bibitem[\protect\citeauthoryear{{Bose} et~al.,}{{Bose}
  et~al.}{2015b}]{Bose2015b}
{Bose} S.,  et~al., 2015b, \mn@doi [\apj] {10.1088/0004-637X/806/2/160}, \href
  {http://adsabs.harvard.edu/abs/2015ApJ...806..160B} {806, 160}

\bibitem[\protect\citeauthoryear{{Cardelli}, {Clayton}  \& {Mathis}}{{Cardelli}
  et~al.}{1989}]{Cardelli1989}
{Cardelli} J.~A.,  {Clayton} G.~C.,   {Mathis} J.~S.,  1989, \mn@doi [\apj]
  {10.1086/167900}, \href {http://adsabs.harvard.edu/abs/1989ApJ...345..245C}
  {345, 245}

\bibitem[\protect\citeauthoryear{{Dall'Ora} et~al.,}{{Dall'Ora}
  et~al.}{2014}]{DallOra2014}
{Dall'Ora} M.,  et~al., 2014, \mn@doi [\apj] {10.1088/0004-637X/787/2/139},
  \href {http://adsabs.harvard.edu/abs/2014ApJ...787..139D} {787, 139}

\bibitem[\protect\citeauthoryear{{Eastman}, {Schmidt}  \& {Kirshner}}{{Eastman}
  et~al.}{1996}]{Eastman1996}
{Eastman} R.~G.,  {Schmidt} B.~P.,   {Kirshner} R.,  1996, \mn@doi [\apj]
  {10.1086/177563}, \href {http://adsabs.harvard.edu/abs/1996ApJ...466..911E}
  {466, 911}

\bibitem[\protect\citeauthoryear{{Falk} \& {Arnett}}{{Falk} \&
  {Arnett}}{1977}]{FalkArnett1977}
{Falk} S.~W.,  {Arnett} W.~D.,  1977, \mn@doi [\apjs] {10.1086/190440}, \href
  {http://adsabs.harvard.edu/abs/1977ApJS...33..515F} {33, 515}

\bibitem[\protect\citeauthoryear{{Faran} et~al.,}{{Faran}
  et~al.}{2014a}]{Faran2014a}
{Faran} T.,  et~al., 2014a, \mn@doi [\mnras] {10.1093/mnras/stu955}, \href
  {http://adsabs.harvard.edu/abs/2014MNRAS.442..844F} {442, 844}

\bibitem[\protect\citeauthoryear{{Faran} et~al.,}{{Faran}
  et~al.}{2014b}]{Faran2014b}
{Faran} T.,  et~al., 2014b, \mn@doi [\mnras] {10.1093/mnras/stu1760}, \href
  {http://adsabs.harvard.edu/abs/2014MNRAS.445..554F} {445, 554}

\bibitem[\protect\citeauthoryear{{Fraser} et~al.,}{{Fraser}
  et~al.}{2011}]{Fraser2011}
{Fraser} M.,  et~al., 2011, \mn@doi [\mnras]
  {10.1111/j.1365-2966.2011.19370.x}, \href
  {http://adsabs.harvard.edu/abs/2011MNRAS.417.1417F} {417, 1417}

\bibitem[\protect\citeauthoryear{{Gurugubelli}, {Sahu}, {Anupama}  \&
  {Chakradhari}}{{Gurugubelli} et~al.}{2008}]{Gurugubelli2008}
{Gurugubelli} U.~K.,  {Sahu} D.~K.,  {Anupama} G.~C.,   {Chakradhari} N.~K.,
  2008, Bulletin of the Astronomical Society of India, \href
  {http://adsabs.harvard.edu/abs/2008BASI...36...79G} {36, 79}

\bibitem[\protect\citeauthoryear{{Hendry} et~al.,}{{Hendry}
  et~al.}{2006}]{Hendry2006}
{Hendry} M.~A.,  et~al., 2006, \mn@doi [\mnras]
  {10.1111/j.1365-2966.2006.10374.x}, \href
  {http://adsabs.harvard.edu/abs/2006MNRAS.369.1303H} {369, 1303}

\bibitem[\protect\citeauthoryear{{Inserra} et~al.,}{{Inserra}
  et~al.}{2011}]{Inserra2011}
{Inserra} C.,  et~al., 2011, \mn@doi [\mnras]
  {10.1111/j.1365-2966.2011.19128.x}, \href
  {http://adsabs.harvard.edu/abs/2011MNRAS.417..261I} {417, 261}

\bibitem[\protect\citeauthoryear{{Inserra} et~al.,}{{Inserra}
  et~al.}{2012}]{Inserra2012a}
{Inserra} C.,  et~al., 2012, \mn@doi [\mnras]
  {10.1111/j.1365-2966.2012.20685.x}, \href
  {http://adsabs.harvard.edu/abs/2012MNRAS.422.1122I} {422, 1122}

\bibitem[\protect\citeauthoryear{{Kasen} \& {Woosley}}{{Kasen} \&
  {Woosley}}{2009}]{Kasen2009}
{Kasen} D.,  {Woosley} S.~E.,  2009, \mn@doi [\apj]
  {10.1088/0004-637X/703/2/2205}, \href
  {http://adsabs.harvard.edu/abs/2009ApJ...703.2205K} {703, 2205}

\bibitem[\protect\citeauthoryear{{Leonard} et~al.,}{{Leonard}
  et~al.}{2002a}]{Leonard2002}
{Leonard} D.~C.,  et~al., 2002a, \mn@doi [\pasp] {10.1086/324785}, \href
  {http://adsabs.harvard.edu/abs/2002PASP..114...35L} {114, 35}

\bibitem[\protect\citeauthoryear{{Leonard} et~al.,}{{Leonard}
  et~al.}{2002b}]{Leonard2002b}
{Leonard} D.~C.,  et~al., 2002b, \mn@doi [\aj] {10.1086/343771}, \href
  {http://adsabs.harvard.edu/abs/2002AJ....124.2490L} {124, 2490}

\bibitem[\protect\citeauthoryear{{Lusk}}{{Lusk}}{2016}]{Lusk2016}
{Lusk} J.~A.,  2016, {SuperBoL: Module for calculating the bolometric
  luminosities of supernovae}, Astrophysics Source Code Library (\mn@eprint
  {ascl} {1609.019})

\bibitem[\protect\citeauthoryear{{Maguire} et~al.,}{{Maguire}
  et~al.}{2010}]{Maguire2010b}
{Maguire} K.,  et~al., 2010, \mn@doi [\mnras]
  {10.1111/j.1365-2966.2010.16332.x}, \href
  {http://adsabs.harvard.edu/abs/2010MNRAS.404..981M} {404, 981}

\bibitem[\protect\citeauthoryear{{Nakar} \& {Sari}}{{Nakar} \&
  {Sari}}{2010}]{NakarSari2010}
{Nakar} E.,  {Sari} R.,  2010, \mn@doi [\apj] {10.1088/0004-637X/725/1/904},
  \href {http://adsabs.harvard.edu/abs/2010ApJ...725..904N} {725, 904}

\bibitem[\protect\citeauthoryear{{Nakar}, {Poznanski}  \& {Katz}}{{Nakar}
  et~al.}{2016}]{Nakar2016}
{Nakar} E.,  {Poznanski} D.,   {Katz} B.,  2016, \mn@doi [\apj]
  {10.3847/0004-637X/823/2/127}, \href
  {http://adsabs.harvard.edu/abs/2016ApJ...823..127N} {823, 127}

\bibitem[\protect\citeauthoryear{{Pastorello} et~al.,}{{Pastorello}
  et~al.}{2009}]{Pastorello2009}
{Pastorello} A.,  et~al., 2009, \mn@doi [\mnras]
  {10.1111/j.1365-2966.2009.14505.x}, \href
  {http://adsabs.harvard.edu/abs/2009MNRAS.394.2266P} {394, 2266}

\bibitem[\protect\citeauthoryear{{Patat}, {Barbon}, {Cappellaro}  \&
  {Turatto}}{{Patat} et~al.}{1994}]{Patat1994}
{Patat} F.,  {Barbon} R.,  {Cappellaro} E.,   {Turatto} M.,  1994, \aap, \href
  {http://adsabs.harvard.edu/abs/1994A%26A...282..731P} {282, 731}

\bibitem[\protect\citeauthoryear{{Pejcha} \& {Prieto}}{{Pejcha} \&
  {Prieto}}{2015}]{Pejcha2015}
{Pejcha} O.,  {Prieto} J.~L.,  2015, \mn@doi [\apj]
  {10.1088/0004-637X/799/2/215}, \href
  {http://adsabs.harvard.edu/abs/2015ApJ...799..215P} {799, 215}

\bibitem[\protect\citeauthoryear{{Pinto} \& {Eastman}}{{Pinto} \&
  {Eastman}}{2000}]{Pinto2000}
{Pinto} P.~A.,  {Eastman} R.~G.,  2000, \mn@doi [\apj] {10.1086/308380}, \href
  {http://adsabs.harvard.edu/abs/2000ApJ...530..757P} {530, 757}

\bibitem[\protect\citeauthoryear{{Quimby}, {Wheeler}, {H{\"o}flich}, {Akerlof},
  {Brown}  \& {Rykoff}}{{Quimby} et~al.}{2007}]{Quimby2007}
{Quimby} R.~M.,  {Wheeler} J.~C.,  {H{\"o}flich} P.,  {Akerlof} C.~W.,  {Brown}
  P.~J.,   {Rykoff} E.~S.,  2007, \mn@doi [\apj] {10.1086/520532}, \href
  {http://adsabs.harvard.edu/abs/2007ApJ...666.1093Q} {666, 1093}

\bibitem[\protect\citeauthoryear{{Rabinak} \& {Waxman}}{{Rabinak} \&
  {Waxman}}{2011}]{Rabinak2011}
{Rabinak} I.,  {Waxman} E.,  2011, \mn@doi [\apj] {10.1088/0004-637X/728/1/63},
  \href {http://adsabs.harvard.edu/abs/2011ApJ...728...63R} {728, 63}

\bibitem[\protect\citeauthoryear{{Richmond}}{{Richmond}}{2014}]{Richmond2014}
{Richmond} M.~W.,  2014, Journal of the American Association of Variable Star
  Observers (JAAVSO), \href {http://adsabs.harvard.edu/abs/2014JAVSO..42..333R}
  {42, 333}

\bibitem[\protect\citeauthoryear{{Roy} \& {Kumar}}{{Roy} \&
  {Kumar}}{2012}]{Roy2012}
{Roy} R.,  {Kumar} B.,  2012, in Astronomical Society of India Conference
  Series. p.~115

\bibitem[\protect\citeauthoryear{{Shussman}, {Waldman}  \& {Nakar}}{{Shussman}
  et~al.}{2016a}]{Shussman2016b}
{Shussman} T.,  {Waldman} R.,   {Nakar} E.,  2016a, preprint, \href
  {http://adsabs.harvard.edu/abs/2016arXiv161005323S} {} (\mn@eprint {arXiv}
  {1610.05323})

\bibitem[\protect\citeauthoryear{{Shussman}, {Nakar}, {Waldman}  \&
  {Katz}}{{Shussman} et~al.}{2016b}]{Shussman2016a}
{Shussman} T.,  {Nakar} E.,  {Waldman} R.,   {Katz} B.,  2016b, preprint, \href
  {http://adsabs.harvard.edu/abs/2016arXiv160202774S} {} (\mn@eprint {arXiv}
  {1602.02774})

\bibitem[\protect\citeauthoryear{{Smartt}}{{Smartt}}{2015}]{Smartt2015}
{Smartt} S.~J.,  2015, \mn@doi [\pasa] {10.1017/pasa.2015.17}, \href
  {http://adsabs.harvard.edu/abs/2015PASA...32...16S} {32, e016}

\bibitem[\protect\citeauthoryear{{Tak{\'a}ts} et~al.,}{{Tak{\'a}ts}
  et~al.}{2014}]{Takats2014}
{Tak{\'a}ts} K.,  et~al., 2014, \mn@doi [\mnras] {10.1093/mnras/stt2203}, \href
  {http://adsabs.harvard.edu/abs/2014MNRAS.438..368T} {438, 368}

\bibitem[\protect\citeauthoryear{{Tak{\'a}ts} et~al.,}{{Tak{\'a}ts}
  et~al.}{2015}]{Takats2015}
{Tak{\'a}ts} K.,  et~al., 2015, \mn@doi [\mnras] {10.1093/mnras/stv857}, \href
  {http://adsabs.harvard.edu/abs/2015MNRAS.450.3137T} {450, 3137}

\bibitem[\protect\citeauthoryear{{Tomasella} et~al.,}{{Tomasella}
  et~al.}{2013}]{Tomasella2013}
{Tomasella} L.,  et~al., 2013, \mn@doi [\mnras] {10.1093/mnras/stt1130}, \href
  {http://adsabs.harvard.edu/abs/2013MNRAS.434.1636T} {434, 1636}

\bibitem[\protect\citeauthoryear{{Tominaga}, {Morokuma}, {Blinnikov},
  {Baklanov}, {Sorokina}  \& {Nomoto}}{{Tominaga} et~al.}{2011}]{Tominaga2011}
{Tominaga} N.,  {Morokuma} T.,  {Blinnikov} S.~I.,  {Baklanov} P.,  {Sorokina}
  E.~I.,   {Nomoto} K.,  2011, \mn@doi [\apjs] {10.1088/0067-0049/193/1/20},
  \href {http://adsabs.harvard.edu/abs/2011ApJS..193...20T} {193, 20}

\bibitem[\protect\citeauthoryear{{Utrobin}}{{Utrobin}}{2007}]{Utrobin2007}
{Utrobin} V.~P.,  2007, \mn@doi [\aap] {10.1051/0004-6361:20066078}, \href
  {http://adsabs.harvard.edu/abs/2007A%26A...461..233U} {461, 233}

\bibitem[\protect\citeauthoryear{{Valenti} et~al.,}{{Valenti}
  et~al.}{2014}]{Valenti2014}
{Valenti} S.,  et~al., 2014, \mn@doi [\mnras] {10.1093/mnrasl/slt171}, \href
  {http://adsabs.harvard.edu/abs/2014MNRAS.438L.101V} {438, L101}

\bibitem[\protect\citeauthoryear{{Valenti} et~al.,}{{Valenti}
  et~al.}{2015}]{Valenti2015}
{Valenti} S.,  et~al., 2015, \mn@doi [\mnras] {10.1093/mnras/stv208}, \href
  {http://adsabs.harvard.edu/abs/2015MNRAS.448.2608V} {448, 2608}

\bibitem[\protect\citeauthoryear{{Valenti} et~al.,}{{Valenti}
  et~al.}{2016}]{Valenti2016}
{Valenti} S.,  et~al., 2016, \mn@doi [\mnras] {10.1093/mnras/stw870}, \href
  {http://adsabs.harvard.edu/abs/2016MNRAS.459.3939V} {459, 3939}

\bibitem[\protect\citeauthoryear{{Van Dyk}, {Li}  \& {Filippenko}}{{Van Dyk}
  et~al.}{2003a}]{VanDyk2003a}
{Van Dyk} S.~D.,  {Li} W.,   {Filippenko} A.~V.,  2003a, \mn@doi [\pasp]
  {10.1086/374299}, \href {http://adsabs.harvard.edu/abs/2003PASP..115..448V}
  {115, 448}

\bibitem[\protect\citeauthoryear{{Van Dyk}, {Li}  \& {Filippenko}}{{Van Dyk}
  et~al.}{2003b}]{VanDyk2003b}
{Van Dyk} S.~D.,  {Li} W.,   {Filippenko} A.~V.,  2003b, \mn@doi [\pasp]
  {10.1086/378308}, \href {http://adsabs.harvard.edu/abs/2003PASP..115.1289V}
  {115, 1289}

\bibitem[\protect\citeauthoryear{{Van Dyk} et~al.,}{{Van Dyk}
  et~al.}{2012}]{VanDyk2012}
{Van Dyk} S.~D.,  et~al., 2012, \mn@doi [\apj] {10.1088/0004-637X/756/2/131},
  \href {http://adsabs.harvard.edu/abs/2012ApJ...756..131V} {756, 131}

\bibitem[\protect\citeauthoryear{{Young}}{{Young}}{2004}]{Young2004}
{Young} T.~R.,  2004, \mn@doi [\apj] {10.1086/425675}, \href
  {http://adsabs.harvard.edu/abs/2004ApJ...617.1233Y} {617, 1233}

\makeatother
\end{thebibliography}


\appendix

\section{A List of the Temperatures and Bolometric Luminosities }
\label{s:results}

\begin{table*}
\caption{A List of the Temperatures and Bolometric Luminosities}
\begin{center}
\def\arraystretch{1.5}
\begin{tabular}{p{2cm}p{1.5cm}p{2cm}p{2cm}}
\hline
MJD & $t-t_\textrm{explosion}$ & T(K) & L(10$^{42}$ erg $\cdot s^{-1}$) \\ 
\hline
\multicolumn{4}{l}{\bf{SN1999em}}\\ 
\hline
51481 & 5 & 11670$^{904}_{761}$ & 1.43$^{0.19}_{0.14}$ \\ 
51482 & 6 & 11748$^{841}_{707}$ & 1.53$^{0.17}_{0.15}$ \\ 
51483 & 7 & 11493$^{652}_{597}$ & 1.50$^{0.14}_{0.12}$ \\ 
51484 & 8 & 11075$^{632}_{532}$ & 1.42$^{0.13}_{0.10}$ \\ 
51485 & 9 & 10681$^{576}_{529}$ & 1.35$^{0.11}_{0.09}$ \\ 
51486 & 10 & 10320$^{506}_{441}$ & 1.27$^{0.09}_{0.07}$ \\
\end{tabular}
\end{center}
\end{table*}
\begin{table*}
\begin{center}
\contcaption{A List of the Temperatures and Bolometric Luminosities}\def\arraystretch{1.5}
\begin{tabular}{p{2cm}p{1.5cm}p{2cm}p{1.6cm}}
51558 & 82 & 5891$^{341}_{324}$ & 0.76$^{0.09}_{0.08}$ \\ 
51565 & 89 & 6080$^{166}_{169}$ & 0.77$^{0.04}_{0.04}$ \\ 
51570 & 94 & 6200$^{164}_{172}$ & 0.76$^{0.03}_{0.04}$ \\ 
51572 & 96 & 6200$^{165}_{171}$ & 0.74$^{0.03}_{0.03}$ \\ 
51576 & 100 & 6480$^{188}_{183}$ & 0.72$^{0.04}_{0.03}$ \\ 
51578 & 102 & 6600$^{193}_{193}$ & 0.71$^{0.04}_{0.04}$ \\ 
51581 & 105 & 6780$^{280}_{266}$ & 0.69$^{0.04}_{0.04}$ \\ 
51586 & 110 & 5400$^{538}_{449}$ & 0.40$^{0.08}_{0.06}$ \\
\end{tabular}
\end{center}
\end{table*}
\begin{table*}
\begin{center}
\contcaption{A List of the Temperatures and Bolometric Luminosities}
\def\arraystretch{1.5}
\begin{tabular}{p{2cm}p{1.5cm}p{2cm}p{1.6cm}}
51550 & 31 & 6120$^{334}_{292}$ & 0.52$^{0.02}_{0.02}$ \\ 
51556 & 37 & 5920$^{308}_{274}$ & 0.50$^{0.01}_{0.01}$ \\ 
51577 & 58 & 5320$^{243}_{219}$ & 0.50$^{0.01}_{0.01}$ \\ 
51582 & 63 & 5260$^{232}_{219}$ & 0.49$^{0.01}_{0.01}$ \\ 
51590 & 71 & 5080$^{223}_{197}$ & 0.50$^{0.02}_{0.02}$ \\ 
51614 & 95 & 5140$^{218}_{211}$ & 0.45$^{0.02}_{0.01}$ \\ 
51619 & 100 & 4960$^{200}_{198}$ & 0.45$^{0.02}_{0.01}$ \\ 
51624 & 105 & 4880$^{205}_{181}$ & 0.43$^{0.02}_{0.01}$ \\
\end{tabular}
\end{center}
\end{table*}
\begin{table*}
\begin{center}
\contcaption{A List of the Temperatures and Bolometric Luminosities}
\def\arraystretch{1.5}
\begin{tabular}{p{2cm}p{1.5cm}p{2cm}p{1.6cm}}
52100 & 14 & 9878$^{1460}_{522}$ & 2.48$^{0.63}_{0.18}$ \\ 
52102 & 16 & 10020$^{1028}_{853}$ & 2.49$^{0.42}_{0.30}$ \\ 
52104 & 18 & 9680$^{954}_{787}$ & 2.30$^{0.36}_{0.26}$ \\ 
52106 & 20 & 9460$^{1017}_{812}$ & 2.17$^{0.37}_{0.25}$ \\ 
52112 & 26 & 8400$^{699}_{568}$ & 1.65$^{0.18}_{0.12}$ \\ 
52117 & 31 & 7740$^{553}_{495}$ & 1.38$^{0.11}_{0.09}$ \\ 
52122 & 36 & 7260$^{500}_{422}$ & 1.22$^{0.08}_{0.06}$ \\ 
52132 & 46 & 6660$^{402}_{356}$ & 1.06$^{0.05}_{0.04}$ \\
\end{tabular}
\end{center}
\end{table*}
\begin{table*}
\begin{center}
\contcaption{A List of the Temperatures and Bolometric Luminosities}
\def\arraystretch{1.5}
\begin{tabular}{p{2cm}p{1.5cm}p{2cm}p{1.6cm}}
\hline
\multicolumn{4}{l}{\bf{SN2001fa}}\\ 
\hline
52202 & 4 & 14820$^{1262}_{1073}$ & 12.54$^{2.21}_{1.66}$ \\ 
52203 & 5 & 13500$^{1011}_{879}$ & 11.47$^{1.65}_{1.29}$ \\ 
52204 & 6 & 11980$^{774}_{671}$ & 9.75$^{1.09}_{0.86}$ \\ 
52205 & 7 & 11320$^{682}_{591}$ & 9.22$^{0.90}_{0.72}$ \\ 
52206 & 8 & 10580$^{583}_{510}$ & 8.46$^{0.70}_{0.56}$ \\ 
52207 & 9 & 9800$^{479}_{441}$ & 7.38$^{0.49}_{0.42}$ \\ 
52208 & 10 & 9560$^{458}_{414}$ & 6.97$^{0.44}_{0.37}$ \\
\end{tabular}
\end{center}
\end{table*}
\begin{table*}
\begin{center}
\contcaption{A List of the Temperatures and Bolometric Luminosities}
\def\arraystretch{1.5}
\begin{tabular}{p{2cm}p{1.5cm}p{2cm}p{1.6cm}}
52103 & 140 & 4500$^{164}_{161}$ & 0.19$^{0.01}_{0.01}$ \\ 
52110 & 147 & 4580$^{173}_{164}$ & 0.18$^{0.01}_{0.01}$ \\ 
\hline
\multicolumn{4}{l}{\bf{SN2002gd}}\\ 
\hline
52555 & 2 & 13120$^{985}_{826}$ & 1.04$^{0.15}_{0.11}$ \\ 
52556 & 3 & 12340$^{839}_{733}$ & 1.02$^{0.12}_{0.10}$ \\ 
52559 & 6 & 10080$^{522}_{477}$ & 0.99$^{0.07}_{0.06}$ \\ 
52562 & 9 & 8640$^{441}_{378}$ & 0.77$^{0.05}_{0.04}$ \\ 
52570 & 17 & 8580$^{1633}_{1118}$ & 0.74$^{0.23}_{0.12}$ \\
\end{tabular}
\end{center}
\end{table*}
\begin{table*}
\begin{center}
\contcaption{A List of the Temperatures and Bolometric Luminosities}
\def\arraystretch{1.5}
\begin{tabular}{p{2cm}p{1.5cm}p{2cm}p{1.6cm}}
52898 & 34 & 6980$^{444}_{394}$ & 3.25$^{0.18}_{0.14}$ \\ 
52900 & 36 & 6700$^{459}_{400}$ & 3.00$^{0.15}_{0.12}$ \\ 
52902 & 38 & 6505$^{965}_{560}$ & 2.78$^{0.41}_{0.17}$ \\ 
52904 & 40 & 6660$^{945}_{729}$ & 2.70$^{0.38}_{0.21}$ \\ 
52906 & 42 & 6740$^{1007}_{765}$ & 2.60$^{0.39}_{0.22}$ \\ 
52910 & 46 & 6380$^{896}_{671}$ & 2.40$^{0.29}_{0.16}$ \\ 
52913 & 49 & 6220$^{936}_{723}$ & 2.31$^{0.28}_{0.16}$ \\ 
52919 & 55 & 6780$^{1205}_{865}$ & 2.07$^{0.40}_{0.21}$ \\
\end{tabular}
\end{center}
\end{table*}
\begin{table*}
\begin{center}
\contcaption{A List of the Temperatures and Bolometric Luminosities}
\def\arraystretch{1.5}
\begin{tabular}{p{2cm}p{1.5cm}p{2cm}p{1.6cm}}
52960 & 40 & 6280$^{430}_{377}$ & 1.20$^{0.04}_{0.04}$ \\ 
52963 & 43 & 6860$^{437}_{372}$ & 1.19$^{0.06}_{0.05}$ \\ 
52966 & 46 & 6520$^{373}_{348}$ & 1.15$^{0.05}_{0.04}$ \\ 
52969 & 49 & 6540$^{398}_{332}$ & 1.13$^{0.05}_{0.04}$ \\ 
52972 & 52 & 6240$^{352}_{305}$ & 1.13$^{0.04}_{0.03}$ \\ 
52977 & 57 & 5560$^{329}_{284}$ & 1.15$^{0.04}_{0.03}$ \\ 
52992 & 72 & 6000$^{309}_{292}$ & 0.99$^{0.03}_{0.03}$ \\ 
52996 & 76 & 5980$^{324}_{276}$ & 0.94$^{0.03}_{0.03}$ \\
\end{tabular}
\end{center}
\end{table*}
\begin{table*}
\begin{center}
\contcaption{A List of the Temperatures and Bolometric Luminosities}
\def\arraystretch{1.5}
\begin{tabular}{p{2cm}p{1.5cm}p{2cm}p{1.6cm}}
53065 & 58 & 5740$^{309}_{282}$ & 0.90$^{0.02}_{0.02}$ \\ 
53066 & 59 & 5760$^{320}_{276}$ & 0.91$^{0.02}_{0.02}$ \\ 
53076 & 69 & 5700$^{300}_{281}$ & 0.91$^{0.02}_{0.02}$ \\ 
53081 & 74 & 5710$^{309}_{277}$ & 0.90$^{0.03}_{0.03}$ \\ 
53099 & 92 & 5620$^{302}_{264}$ & 0.82$^{0.02}_{0.02}$ \\ 
\hline
\multicolumn{4}{l}{\bf{SN2004du}}\\ 
\hline
53231 & 3 & 14644$^{2903}_{1658}$ & 4.53$^{1.64}_{0.75}$ \\ 
53233 & 5 & 11420$^{700}_{620}$ & 3.61$^{0.36}_{0.30}$ \\
\end{tabular}
\end{center}
\end{table*}
\begin{table*}
\begin{center}
\contcaption{A List of the Temperatures and Bolometric Luminosities}
\def\arraystretch{1.5}
\begin{tabular}{p{2cm}p{1.5cm}p{2cm}p{1.6cm}}
53286 & 15 & 9750$^{538}_{473}$ & 2.00$^{0.15}_{0.12}$ \\ 
53287 & 16 & 9680$^{507}_{462}$ & 1.98$^{0.14}_{0.12}$ \\ 
53288 & 17 & 9600$^{506}_{447}$ & 1.97$^{0.14}_{0.11}$ \\ 
53289 & 18 & 9500$^{500}_{432}$ & 1.96$^{0.14}_{0.11}$ \\ 
53294 & 23 & 8300$^{201}_{195}$ & 1.65$^{0.06}_{0.06}$ \\ 
53296 & 25 & 8021$^{411}_{235}$ & 1.57$^{0.14}_{0.07}$ \\ 
53297 & 26 & 7989$^{306}_{264}$ & 1.58$^{0.10}_{0.08}$ \\ 
53298 & 27 & 7880$^{248}_{235}$ & 1.55$^{0.08}_{0.07}$ \\
\end{tabular}
\end{center}
\end{table*}
\begin{table*}
\begin{center}
\contcaption{A List of the Temperatures and Bolometric Luminosities}
\def\arraystretch{1.5}
\begin{tabular}{p{2cm}p{1.5cm}p{2cm}p{1.6cm}}
53554 & 5 & 13509$^{693}_{623}$ & 0.63$^{0.06}_{0.05}$ \\ 
53555 & 6 & 12356$^{846}_{692}$ & 0.54$^{0.06}_{0.04}$ \\ 
53557 & 8 & 10418$^{788}_{668}$ & 0.42$^{0.03}_{0.03}$ \\ 
53558 & 9 & 9690$^{317}_{298}$ & 0.38$^{0.02}_{0.02}$ \\ 
53559 & 10 & 9540$^{295}_{271}$ & 0.38$^{0.02}_{0.02}$ \\ 
53560 & 11 & 9269$^{374}_{364}$ & 0.36$^{0.02}_{0.02}$ \\ 
53562 & 13 & 8376$^{468}_{333}$ & 0.31$^{0.03}_{0.02}$ \\ 
53563 & 14 & 8088$^{246}_{230}$ & 0.30$^{0.02}_{0.02}$ \\
\end{tabular}
\end{center}
\end{table*}
\begin{table*}
\begin{center}
\contcaption{A List of the Temperatures and Bolometric Luminosities}
\def\arraystretch{1.5}
\begin{tabular}{p{2cm}p{1.5cm}p{2cm}p{1.6cm}}
53642 & 93 & 5520$^{583}_{460}$ & 0.24$^{0.02}_{0.01}$ \\ 
53781 & 232 & 4080$^{338}_{283}$ & 0.01$^{0.01}_{0.00}$ \\ 
\hline
\multicolumn{4}{l}{\bf{SN2006bp}}\\ 
\hline
53836 & 2 & 10662$^{480}_{632}$ & 0.90$^{0.06}_{0.06}$ \\ 
53837 & 3 & 9779$^{267}_{208}$ & 0.91$^{0.04}_{0.04}$ \\ 
53838 & 4 & 9380$^{163}_{152}$ & 0.95$^{0.02}_{0.02}$ \\ 
53839 & 5 & 8713$^{171}_{215}$ & 0.97$^{0.02}_{0.02}$ \\ 
53840 & 6 & 8292$^{225}_{173}$ & 0.97$^{0.02}_{0.02}$ \\
\end{tabular}
\end{center}
\end{table*}
\begin{table*}
\begin{center}
\contcaption{A List of the Temperatures and Bolometric Luminosities}
\def\arraystretch{1.5}
\begin{tabular}{p{2cm}p{1.5cm}p{2cm}p{1.6cm}}
54436 & 37 & 7420$^{201}_{192}$ & 4.46$^{0.22}_{0.20}$ \\ 
54439 & 40 & 7150$^{245}_{232}$ & 4.12$^{0.27}_{0.23}$ \\ 
54441 & 42 & 7060$^{161}_{168}$ & 3.99$^{0.15}_{0.16}$ \\ 
54442 & 43 & 7000$^{145}_{154}$ & 3.91$^{0.13}_{0.14}$ \\ 
54446 & 47 & 6860$^{137}_{148}$ & 3.70$^{0.12}_{0.12}$ \\ 
54449 & 50 & 6690$^{149}_{146}$ & 3.46$^{0.13}_{0.13}$ \\ 
54450 & 51 & 6800$^{146}_{152}$ & 3.54$^{0.13}_{0.13}$ \\ 
54451 & 52 & 6800$^{149}_{155}$ & 3.54$^{0.13}_{0.13}$ \\
\end{tabular}
\end{center}
\end{table*}
\begin{table*}
\begin{center}
\contcaption{A List of the Temperatures and Bolometric Luminosities}
\def\arraystretch{1.5}
\begin{tabular}{p{2cm}p{1.5cm}p{2cm}p{1.6cm}}
54900 & 78 & 5940$^{286}_{271}$ & 0.29$^{0.01}_{0.01}$ \\ 
54901 & 79 & 6460$^{679}_{542}$ & 0.31$^{0.03}_{0.02}$ \\ 
54906 & 84 & 5880$^{257}_{246}$ & 0.28$^{0.01}_{0.01}$ \\ 
54912 & 90 & 6000$^{335}_{288}$ & 0.26$^{0.01}_{0.01}$ \\ 
54915 & 93 & 5840$^{193}_{187}$ & 0.24$^{0.01}_{0.01}$ \\ 
54917 & 95 & 5860$^{193}_{167}$ & 0.23$^{0.01}_{0.01}$ \\ 
54925 & 103 & 5960$^{248}_{225}$ & 0.20$^{0.01}_{0.01}$ \\ 
54927 & 105 & 5649$^{250}_{213}$ & 0.17$^{0.01}_{0.00}$ \\
\end{tabular}
\end{center}
\end{table*}
\begin{table*}
\begin{center}
\contcaption{A List of the Temperatures and Bolometric Luminosities}
\def\arraystretch{1.5}
\begin{tabular}{p{2cm}p{1.5cm}p{2cm}p{1.6cm}}
54918 & 73 & 5280$^{120}_{101}$ & 0.53$^{0.01}_{0.01}$ \\ 
54919 & 74 & 5280$^{117}_{105}$ & 0.53$^{0.01}_{0.01}$ \\ 
54920 & 75 & 5280$^{115}_{105}$ & 0.53$^{0.01}_{0.01}$ \\ 
54946 & 101 & 5300$^{113}_{109}$ & 0.39$^{0.01}_{0.01}$ \\ 
54947 & 102 & 5060$^{109}_{122}$ & 0.39$^{0.01}_{0.01}$ \\ 
\hline
\multicolumn{4}{l}{\bf{SN2009bw}}\\ 
\hline
54923 & 6 & 15826$^{988}_{757}$ & 2.29$^{0.64}_{0.09}$ \\ 
54924 & 7 & 14043$^{365}_{356}$ & 2.16$^{0.08}_{0.08}$ \\
\end{tabular}
\end{center}
\end{table*}
\begin{table*}
\begin{center}
\contcaption{A List of the Temperatures and Bolometric Luminosities}
\def\arraystretch{1.5}
\begin{tabular}{p{2cm}p{1.5cm}p{2cm}p{1.6cm}}
55063 & 22 & 7860$^{626}_{519}$ & 0.65$^{0.10}_{0.07}$ \\ 
55064 & 23 & 7760$^{590}_{496}$ & 0.64$^{0.07}_{0.05}$ \\ 
55066 & 25 & 7330$^{543}_{444}$ & 0.60$^{0.07}_{0.04}$ \\ 
55067 & 26 & 7340$^{512}_{453}$ & 0.60$^{0.05}_{0.04}$ \\ 
55068 & 27 & 7109$^{509}_{409}$ & 0.57$^{0.05}_{0.04}$ \\ 
55069 & 28 & 7150$^{479}_{410}$ & 0.59$^{0.04}_{0.04}$ \\ 
55070 & 29 & 6920$^{420}_{379}$ & 0.55$^{0.04}_{0.03}$ \\ 
55071 & 30 & 6840$^{415}_{365}$ & 0.55$^{0.04}_{0.03}$ \\
\end{tabular}
\end{center}
\end{table*}
\begin{table*}
\begin{center}
\contcaption{A List of the Temperatures and Bolometric Luminosities}
\def\arraystretch{1.5}
\begin{tabular}{p{2cm}p{1.5cm}p{2cm}p{1.6cm}}
55137 & 96 & 5480$^{237}_{219}$ & 0.48$^{0.01}_{0.01}$ \\ 
55138 & 97 & 5460$^{239}_{214}$ & 0.48$^{0.01}_{0.01}$ \\ 
55141 & 100 & 5520$^{259}_{230}$ & 0.48$^{0.01}_{0.01}$ \\ 
55146 & 105 & 5460$^{242}_{209}$ & 0.47$^{0.01}_{0.01}$ \\ 
55147 & 106 & 5460$^{237}_{213}$ & 0.47$^{0.01}_{0.01}$ \\ 
\hline
\multicolumn{4}{l}{\bf{SN2012A}}\\ 
\hline
55937 & 8 & 15960$^{636}_{594}$ & 1.62$^{0.12}_{0.11}$ \\ 
55938 & 9 & 13700$^{1166}_{641}$ & 1.43$^{0.21}_{0.14}$ \\
\end{tabular}
\end{center}
\end{table*}
\begin{table*}
\begin{center}
\def\arraystretch{1.5}
\contcaption{A List of the Temperatures and Bolometric Luminosities}
\begin{tabular}{p{2cm}p{1.5cm}p{2cm}p{1.6cm}}
55999 & 70 & 6653$^{419}_{350}$ & 0.46$^{0.03}_{0.03}$ \\ 
56000 & 71 & 6580$^{361}_{320}$ & 0.46$^{0.02}_{0.02}$ \\ 
56001 & 72 & 6665$^{361}_{328}$ & 0.45$^{0.03}_{0.02}$ \\ 
56002 & 73 & 6657$^{431}_{340}$ & 0.43$^{0.03}_{0.02}$ \\ 
56003 & 74 & 6600$^{328}_{289}$ & 0.43$^{0.02}_{0.02}$ \\ 
56004 & 75 & 6600$^{329}_{306}$ & 0.43$^{0.02}_{0.02}$ \\ 
56005 & 76 & 6620$^{323}_{291}$ & 0.42$^{0.02}_{0.02}$ \\ 
56006 & 77 & 6600$^{280}_{266}$ & 0.42$^{0.02}_{0.02}$ \\
\end{tabular}
\end{center}
\end{table*}
\begin{table*}
\begin{center}
\contcaption{A List of the Temperatures and Bolometric Luminosities}
\def\arraystretch{1.5}
\begin{tabular}{p{2cm}p{1.5cm}p{2cm}p{1.6cm}}
56022 & 20 & 7640$^{124}_{111}$ & 1.49$^{0.04}_{0.04}$ \\ 
56024 & 22 & 7400$^{127}_{124}$ & 1.46$^{0.04}_{0.04}$ \\ 
56025 & 23 & 7377$^{182}_{226}$ & 1.46$^{0.06}_{0.06}$ \\ 
56026 & 24 & 7260$^{159}_{165}$ & 1.43$^{0.06}_{0.05}$ \\ 
56027 & 25 & 7160$^{160}_{150}$ & 1.40$^{0.06}_{0.05}$ \\ 
56028 & 26 & 7120$^{148}_{156}$ & 1.40$^{0.06}_{0.05}$ \\ 
56029 & 27 & 7020$^{162}_{144}$ & 1.37$^{0.05}_{0.05}$ \\ 
56030 & 28 & 6965$^{177}_{180}$ & 1.36$^{0.05}_{0.05}$ \\
\end{tabular}
\end{center}
\end{table*}
\begin{table*}
\begin{center}
\contcaption{A List of the Temperatures and Bolometric Luminosities}
\def\arraystretch{1.5}
\begin{tabular}{p{2cm}p{1.5cm}p{2cm}p{1.6cm}}
\hline
\multicolumn{4}{l}{\bf{SN2012ec}}\\ 
\hline
56154 & 11 & 10020$^{345}_{326}$ & 2.26$^{0.09}_{0.08}$ \\ 
56155 & 12 & 9780$^{332}_{304}$ & 2.14$^{0.08}_{0.07}$ \\ 
56158 & 15 & 9068$^{926}_{534}$ & 1.85$^{0.16}_{0.13}$ \\ 
56159 & 16 & 8080$^{203}_{196}$ & 1.69$^{0.04}_{0.04}$ \\ 
56165 & 22 & 6880$^{101}_{107}$ & 1.44$^{0.15}_{0.13}$ \\ 
56168 & 25 & 6640$^{101}_{88}$ & 1.36$^{0.14}_{0.12}$ \\ 
56171 & 28 & 6480$^{92}_{100}$ & 1.33$^{0.13}_{0.10}$ \\
\end{tabular}
\end{center}
\end{table*}
\begin{table*}
\begin{center}
\contcaption{A List of the Temperatures and Bolometric Luminosities}
\def\arraystretch{1.5}
\begin{tabular}{p{2cm}p{1.5cm}p{2cm}p{1.6cm}}
56363 & 23 & 7920$^{223}_{202}$ & 1.74$^{0.04}_{0.04}$ \\ 
56364 & 24 & 7560$^{202}_{182}$ & 1.67$^{0.04}_{0.03}$ \\ 
56365 & 25 & 7540$^{195}_{186}$ & 1.65$^{0.03}_{0.03}$ \\ 
56366 & 26 & 7380$^{215}_{198}$ & 1.57$^{0.04}_{0.04}$ \\ 
56368 & 28 & 7190$^{560}_{206}$ & 1.51$^{0.10}_{0.04}$ \\ 
56370 & 30 & 7080$^{354}_{335}$ & 1.44$^{0.06}_{0.05}$ \\ 
56372 & 32 & 7016$^{452}_{380}$ & 1.40$^{0.08}_{0.05}$ \\ 
56373 & 33 & 6940$^{349}_{309}$ & 1.37$^{0.05}_{0.04}$ \\
\end{tabular}
\end{center}
\end{table*}
\begin{table*}
\begin{center}
\contcaption{A List of the Temperatures and Bolometric Luminosities}
\def\arraystretch{1.5}
\begin{tabular}{p{2cm}p{1.5cm}p{2cm}p{1.6cm}}
56429 & 89 & 6020$^{272}_{247}$ & 0.95$^{0.02}_{0.02}$ \\ 
56430 & 90 & 5860$^{255}_{223}$ & 0.92$^{0.02}_{0.02}$ \\ 
56431 & 91 & 5760$^{246}_{215}$ & 0.90$^{0.02}_{0.02}$ \\ 
56432 & 92 & 6180$^{330}_{280}$ & 0.90$^{0.03}_{0.02}$ \\ 
56433 & 93 & 5776$^{646}_{393}$ & 0.86$^{0.06}_{0.03}$ \\ 
56434 & 94 & 5580$^{229}_{207}$ & 0.83$^{0.02}_{0.02}$ \\ 
56435 & 95 & 5663$^{444}_{302}$ & 0.82$^{0.02}_{0.02}$ \\ 
56436 & 96 & 5900$^{299}_{254}$ & 0.81$^{0.02}_{0.02}$ \\
\end{tabular}
\end{center}
\end{table*}
\begin{table*}
\begin{center}
\contcaption{A List of the Temperatures and Bolometric Luminosities}
\def\arraystretch{1.5}
\begin{tabular}{p{2cm}p{1.5cm}p{2cm}p{1.6cm}}
56410 & 3 & 14923$^{704}_{552}$ & 13.53$^{1.09}_{0.79}$ \\ 
56411 & 4 & 13940$^{350}_{343}$ & 12.41$^{0.51}_{0.48}$ \\ 
56412 & 5 & 12422$^{491}_{1173}$ & 10.76$^{0.65}_{1.36}$ \\ 
56413 & 6 & 11304$^{440}_{389}$ & 9.73$^{0.51}_{0.43}$ \\ 
56414 & 7 & 10875$^{562}_{477}$ & 9.09$^{0.70}_{0.48}$ \\ 
56416 & 9 & 9991$^{474}_{437}$ & 8.12$^{0.34}_{0.32}$ \\ 
56417 & 10 & 9500$^{289}_{275}$ & 7.47$^{0.24}_{0.22}$ \\ 
56418 & 11 & 9340$^{287}_{253}$ & 7.29$^{0.23}_{0.19}$ \\
\end{tabular}
\end{center}
\end{table*}
\begin{table*}
\begin{center}
\contcaption{A List of the Temperatures and Bolometric Luminosities}
\def\arraystretch{1.5}
\begin{tabular}{p{2cm}p{1.5cm}p{2cm}p{1.6cm}} 
56481 & 74 & 5183$^{296}_{278}$ & 1.40$^{0.06}_{0.07}$ \\ 
56482 & 75 & 5173$^{286}_{261}$ & 1.34$^{0.05}_{0.05}$ \\ 
56483 & 76 & 5188$^{294}_{258}$ & 1.26$^{0.05}_{0.04}$ \\ 
56486 & 79 & 5080$^{270}_{255}$ & 1.07$^{0.05}_{0.04}$ \\ 
56487 & 80 & 5012$^{282}_{263}$ & 1.00$^{0.05}_{0.05}$ \\ 
56488 & 81 & 4816$^{255}_{234}$ & 0.88$^{0.05}_{0.04}$ \\ 
56489 & 82 & 5172$^{391}_{433}$ & 0.76$^{0.05}_{0.10}$ \\ 
\hline
\multicolumn{4}{l}{\bf{SN2013ej}}\\
\end{tabular}
\end{center}
\end{table*}
\begin{table*}
\begin{center}
\contcaption{A List of the Temperatures and Bolometric Luminosities}
\def\arraystretch{1.5}
\begin{tabular}{p{2cm}p{1.5cm}p{2cm}p{1.6cm}}
56563 & 66 & 6800$^{1097}_{805}$ & 1.34$^{0.24}_{0.12}$ \\ 
56567 & 70 & 6440$^{953}_{732}$ & 1.25$^{0.17}_{0.09}$ \\ 
56568 & 71 & 6340$^{915}_{713}$ & 1.21$^{0.16}_{0.09}$ \\ 
56573 & 76 & 6600$^{1003}_{771}$ & 1.16$^{0.18}_{0.10}$ \\ 
56574 & 77 & 6800$^{1097}_{805}$ & 1.16$^{0.21}_{0.11}$ \\ 
56577 & 80 & 6920$^{1125}_{843}$ & 1.14$^{0.21}_{0.11}$ \\ 
56578 & 81 & 6540$^{981}_{758}$ & 1.05$^{0.15}_{0.09}$ \\ 
56589 & 92 & 6700$^{1056}_{784}$ & 0.79$^{0.13}_{0.07}$ \\ 
56594 & 97 & 6560$^{993}_{759}$ & 0.59$^{0.09}_{0.05}$ \\ 
56596 & 99 & 5980$^{812}_{630}$ & 0.44$^{0.04}_{0.02}$ \\ 
56600 & 103 & 6420$^{1135}_{837}$ & 0.22$^{0.04}_{0.02}$ \\ 
56607 & 110 & 6180$^{1358}_{933}$ & 0.15$^{0.03}_{0.01}$ \\ 
56610 & 113 & 6980$^{2924}_{1539}$ & 0.16$^{0.09}_{0.02}$ \\ 
56613 & 116 & 5960$^{1815}_{1111}$ & 0.12$^{0.03}_{0.01}$ \\ 
56616 & 119 & 6440$^{1638}_{1072}$ & 0.13$^{0.03}_{0.01}$ \\ 
56617 & 120 & 5960$^{1722}_{1075}$ & 0.13$^{0.03}_{0.01}$ \\ 

\hline
\end{tabular}
\end{center}
\end{table*}

\bsp	
\label{lastpage}
\end{document}